\documentclass[preprintnumbers,aps]{revtex4}
\usepackage{amsmath,amssymb,graphicx}
\def\unit{\leavevmode\hbox{\small1\kern-3.6pt\normalsize1}}
 \normalsize

\def\lsim{\raise0.3ex\hbox{$\;<$\kern-0.75em\raise-1.1ex\hbox{$\sim\;$}}}
\def\gsim{\raise0.3ex\hbox{$\;>$\kern-0.75em\raise-1.1ex\hbox{$\sim\;$}}}

\usepackage{graphics}

\newcommand{\be}{\begin{eqnarray}}
\newcommand{\ee}{\end{eqnarray}}

\def\bea{\begin{eqnarray}}
\def\eea{\end{eqnarray}}

\usepackage{epsfig}
 \normalsize

\begin{document}

\title{LHC signals of a BLSSM CP-even Higgs boson}
\author{A. Hammad$^{1}$, S. Khalil$^{1}$, S. Moretti$^{2}$ }
\affiliation{$^{1}$ Center for Fundamental Physics, Zewail City {of} Science and Technology,
6 October  City, Giza, Egypt.\\
$^{2}$ School of Physics \& Astronomy, University of Southampton, Highfield, Southampton, UK.}
\date{\today}

\begin{abstract}
We study the scope of the Large Hadron Collider in accessing a neutral Higgs boson of the 
$B-L$  Supersymmetric Standard Model. After assessing the surviving parameter space configurations following the
Run 1 data taking, we investigate the possibilities of detecting this object during Run 2. For the model configurations
in which the mixing between such a state and the discovered Standard Model-like Higgs boson is non-negligible, there exist
several channels enabling its discovery over a mass range spanning from $\approx 140$ to $\approx$ 500 GeV.
For a heavier Higgs state, with mass above 250 GeV (i.e., twice the 
mass of the Higgs state discovered in 2012), the hallmark signature is its decay in two such 125 GeV scalars, $h'\to hh$,
where $hh\to b\bar b \gamma\gamma$. For a lighter Higgs state, with mass of order 140 GeV, three channels are accessible: $\gamma\gamma$, $Z\gamma$ and 
$ZZ$, wherein the $Z$ boson decays leptonically. In all such cases, significances above discovery can occur for already planned luminosities at the CERN machine.

\end{abstract}
\maketitle
\section{Introduction}
\label{sec:1}

After the Higgs boson discovery at the Large Hadron Collider (LHC)  during Run 1,  a new era in particle physics has begun. While precision measurements of the detected state as reported by the
ATLAS and CMS collaborations (now also including Run 2 data) confirm a Standard Model (SM)-like nature
with a rather light mass of $\approx 125$ GeV, significant effort
is now being put in the search for companion Higgs states, as any Beyond the SM (BSM) construct embedding a Higgs
mechanism is likely being non-minimal, i.e., it would include new Higgs bosons in its spectrum. In the miyriad of
BSM
scenarios  available, a special place is held by models of Supersymmetry (SUSY), wherein the lightest SM-like Higgs boson mass
 is naturally limited to be at the Electro-Weak (EW) scale (say below $2M_W$) and where one also finds additional (neutral) Higgs bosons. Thus, one may well be tempted to conclude that a SUSY scenario may be behind the aforementioned data. 

Amongst the many SUSY realisations studied so far, though, one really ought to single out those that also offer explanations to other data pointing to BSM physics, chiefly those indicating that neutrinos oscillate, hence that they have mass. One is therefore well
 motivated in looking at the $B-L$ Supersymmetric Standard Model (BLSSM). The BLSSM is an extension of the
time-honoured Minimal  Supersymmetric Standard Model (MSSM) obtained by adopting a further $U(1)_{B-L}$ gauge group
alongside the SM structure, i.e., $SU(3)_C \times SU(2)_L \times U(1)_Y \times U(1)_{B-L}$. (This requires
an additional Higgs singlet field to break the new $U(1)_{B-L}$ symmetry, in turn releasing an additional $Z'$ state as well.)
The particle content of the BLSSM,
limited to its Higgs sector, includes three additional neutral Higgs fields (henceforth $h', H'$ and $A'$) with respect to the MSSM ones
(henceforth $h, H$ and $ A$)\footnote{We conventionally denote here by $A^{(')}$ CP-odd Higgs states and
by $h^{(')}, H^{(')}$ CP-even ones (the latter with $m_h< m_H$ and $m_{h'}<m_{H'}$). Notice that also 
two charge conjugated states are  present in both the MSSM and BLSSM, denoted by  $H^\pm$.}.

The enriched Higgs sector of the BLSSM, with respect to the MSSM one, offers the possibility of relieving  the deadlock typical of the minimal SUSY model, wherein a light SM-like Higgs state (the $h$ boson at $\approx 125$ GeV) requires the other Higgs states ($H$
and $A$ in particular) to be much heavier in comparison (and moderately coupled to SM matter fermions and gauge bosons). This does not necessarily occurs in the BLSSM, as the $h', H'$ and $A'$ states can have a singlet component sufficient to render them very lightly mixed with the $h$ one, thereby allowing at the same time sizable couplings to SM objects and the possibility of
their mass, depending on the Vacuum Expectation Value (VEV) of the Higgs singlet field, to be significantly lighter than those of the MSSM-like $H$ and $A$ particles.  

In fact, a natural configuration of the BLSSM is to find   alongside the above 
SM-like Higgs state  another rather light physical Higgs boson, $h'$, also CP-even, with a mass $m_{h'}\geq135$ GeV.
This fact was exploited in Refs.~\cite{Abdallah:2014fra,Khalil:2015vpa,Hammad:2015eca} to explain  potential 
Run 1 signals for another Higgs boson, i.e., $h^\prime$, in the $h^\prime \to ZZ^\ast \to 4l$ (wherein a $2\sigma$ excess is appreciable in the vicinities of 145 GeV \cite{Chatrchyan:2013mxa}),
 $h^\prime \to\gamma\gamma$  (prompting a $2.9\sigma$ excess  around 137 GeV \cite{CERNAA}) and  
$h^\prime \to Z\gamma$ (yielding a $2\sigma$ excess  around $140$ GeV \cite{Chatrchyan:2013vaa})
decay modes. 

As new data are presently being collected at Run 2, we revisit here the scope of the LHC in confirming or disproving the
above hypothesis of additional light Higgs boson signals. Furthermore, thanks to the higher energy and luminosity afforded
by the new CERN machine configuration, we also investigate the possibilities of accessing a  
heavier $h'$ state, with a mass up to 500 GeV or so. Our paper is organised as follows. In the next section, we 
introduce the Higgs boson spectrum in the BLSSM. In Sects.~\ref{sec:heavy} and \ref{sec:light}  we describe our
analysis of the light and heavy, respectively, mass range of the $h'$ state. We conclude in Sect.~\ref{sec:summary}.

\section{Higgs bosons in the BLSSM}
\label{sec:spectrum}

 The BLSSM model consists of, in addition to the MSSM particle content, two SM singlet chiral Higgs superfields $\chi_{1,2}$ and three SM singlet chiral superfields, $\nu_i, i =1,2,3 $ \cite{Khalil:2007dr}. The Superpotential of this model is given by 
\begin{equation} \label{eq:1}
W = Y_u \hat{Q} \hat{H}_u \hat{U}^c +Y_d \hat{Q}\hat{H}_d \hat{D}^c + Y_e \hat{L} \hat{H}_d \hat{E}^c +  Y_\nu \hat{L} \hat{H}_u \hat{\nu}^c \\
+ Y_\nu\,\hat{\nu}\,\hat{l}\,\hat{H}_u. + \mu \hat{H}_u \hat{H}_d + \mu^\prime \hat{\chi}_1\hat{\chi}_2.
\end{equation}
The corresponding soft SUSY breaking terms and the details of the associated spectrum can be found in Refs.~\cite{Khalil:2007dr, Staub:2015kfa}.  Note that the $U(1)_Y$ and $U(1)_{B-L}$ gauge kinetic mixing can be absorbed in the covariant derivative redefinition and, in this basis, one finds
\bea
M_Z^2 &=& \frac{1}{4} (g_1^2 +g_2^2) v^2, \\
M_{Z'}^2 &=& g_{BL}^2 v'^2 + \frac{1}{4} \tilde{g}^2 v^2 ,
\eea
where $g_{BL}$ is the gauge coupling of $U(1)_{B-L}$ and $\tilde{g}$ is the gauge coupling mixing between $U(1)_Y$ and $U(1)_{B-L}$. In addition,  $v=\sqrt{v^2_1+v^2_2}\simeq 246$ GeV,  $v'=\sqrt{v'^2_1+v'^2_2} \simeq {\cal O}(1)$ TeV are the VEVs of the
Higgs fields $H_i$ and $\chi_i$, respectively. 

\subsection{The spectrum}

The neutral Higgs boson masses are obtained by making the usual redefinition of the Higgs fields, i.e.,
$H_{1,2}^0 = {\frac{1}{\sqrt{2}}}(v_{1,2} + \sigma_{1,2} + i \phi_{1,2}) $ and 
$\chi_{1,2}^0 ={\frac{1}{\sqrt{2}}}(v'_{1,2} + \sigma'_{1,2}  + i \phi'_{1,2})$,
where $\sigma_{1,2}= {\rm Re} H_{1,2}^0$, $\phi_{1,2}={\rm Im} H_{1,2}^0$, $\sigma'_{1,2}= {\rm Re} \chi_{1,2}^0$ and $\phi'_{1,2}={\rm Im} \chi_{1,2}^0 $. The real parts correspond to the CP-even Higgs bosons and the imaginary parts correspond to the CP-odd Higgs bosons. Therefore, the squared matrix of  the BLSSM CP-even neutral Higgs fields at tree level, in the basis $(\sigma_1,\sigma_2,\sigma'_1,\sigma'_2)$, is given by
\begin{equation}\label{eq:6}
 M^2 = \left( \begin{array}{cc} 
 			     M^2_{h{H}} ~ &~  M^2_{hh'} \\  \\
                              M^{2^{{T}}}_{hh'} ~ &~  M^2_{h'{H}'} \end{array} \right),
\end{equation}
where $M^2_{h{H}} $ is the usual MSSM neutral CP-even Higgs mass matrix, which leads to a SM-like Higgs boson with mass, at one loop level, of order 125 GeV and a heavy Higgs boson with mass $m_H \sim  {\cal O}(1$ TeV). 
In addition, the BLSSM matrix $M^2_{h'{H}'}$  is given by\\ 
\be
M^2_{h'{H}'}=
\left( \begin{array}{cc}
               m^2_{A'} c^2_{\beta'} + g^2_{BL} v'^2_1 &-\frac{1}{2} m^2_{A'} s_{2\beta'} - g^2_{BL} v'_1 v'_2 \\
                    \\
              -\frac{1}{2} m^2_{A'}s_{2\beta'} - g^2_{BL} v'_1 v'_2 & m^2_{A'} s^2_{\beta'} + g^2_{BL} v'^2_2
              \end{array}\right),
\ee
where $c_x=\cos(x)$ and $s_x=\sin(x)$. Therefore, the eigenvalues of this mass matrix are given by
\bea
{m}^2_{h',H'} = \frac{1}{2} \Big[ ( m^2_{A'} + M_{{Z'}}^2 )
 \mp\sqrt{ ( m^2_{A'} + M_{{Z'}}^2 )^2 - 4 m^2_{A'} M_{{Z'}}^2 \cos^2 2\beta' }\;\Big].
\eea
If $\cos^2{{2}\beta'} \ll 1$, one finds that the lightest $B-L$ neutral Higgs state is given by %
\be%
{m}_{h'}\; {\simeq}\; \left(\frac{m^2_{A'} M_{{Z'}}^2 \cos^2 2\beta'}{{m^2_{A'}+M_{{Z'}}^2}}\right)^{\frac{1}{2}} \simeq {\cal O}(100~ {\rm GeV}).%
\ee%
The mixing matrix $M_{hh'}^2$ is proportional to $\tilde{g}$ and, for a gauge coupling $g_{BL} \sim \vert \tilde{g}\vert  \sim {\cal O}(0.1)$, these off-diagonal terms are about one order of magnitude smaller than the diagonal ones. However, they are still crucial for generating interaction vertices between the light BLSSM Higgs boson, $h'$, and the MSSM-like Higgs state, $h$.

The CP-even neutral Higgs mass matrix in Eq.~(\ref{eq:6}) can be diagonalised by a unitary transformation:
\be 
{\Gamma}~ M^2 ~ \Gamma^\dag = {\rm diag}\{m_h^2, m_{h'}^2, m_H^2, m_{H'}^2\}. 
\ee
A numerical scan confirms that, while $\tan^\prime \beta \le 1.2 $, the $h^\prime $ state can be the second Higgs boson mass whereas the other two CP-even states $H,H^\prime$ are heavy. Also, the mixings $\Gamma_{ij} $ are proportional to $\tilde{g}$ and they vanish (at tree level) if $\tilde{g} = 0$. In this regard, $h'$ can be written in terms of  gauge eigenstates as 
\be 
h' = \Gamma_{21} ~\sigma_1 + \Gamma_{22} ~\sigma_2 + \Gamma_{23}~ \sigma'_1  +\Gamma_{24}~ \sigma'_2.
\ee 
Thus, the couplings of the $h'$ with up- and down-quarks  are given by 
\begin{equation}
g_{h^\prime u \bar{u}} = -i \frac{m_u\times \Gamma_{22}}{\upsilon \sin\beta},
\end{equation}
\begin{equation}
g_{h^\prime d \bar{d}} = -i \frac{m_d\times \Gamma_{21}}{\upsilon \cos\beta}.
\end{equation}
Similarly, one can derive the $h'$ couplings with the ${W^{+}}{W^-}$ and $Z{Z}$ gauge boson pairs:
\bea%
\hspace{-0.65cm} g_{_{h' W W}} &=& {i}~g_{{2}} M_W \left(\Gamma_{{22}} \sin \beta + \Gamma_{{21}} \cos\beta\right),\nonumber\\
g_{_{h' Z Z}} &=&\frac{i}{2}\Big[4 g_{BL}\sin^2{\theta'}\left(v'_1\Gamma_{22} + v'_2\Gamma_{21}\right)+\left(v_2\Gamma_{22} + v_1\Gamma_{21}\right)\left(g_z \cos{\theta'}-\tilde{g}\sin{\theta'}\right)^2\Big],~~
\eea
where $g_z=\sqrt{g_1^2+g_2^2}$ and  $\theta'$ is the mixing angle between $Z$ and $Z'$.  Since $\sin{\theta'}\ll 1$
(as per experimental constraints), the coupling of the $h'$ with $ZZ$, $g_{_{h'ZZ}}$, will be as follows:
\be
g_{_{h'ZZ}}\simeq i~g_z\, M_Z  \left(\Gamma_{22} \sin{\beta}+ \Gamma_{21} \cos{\beta}\right).
\ee

\begin{figure}[ht]
\centering
\includegraphics[height=6.5cm,width=10cm,angle=0]{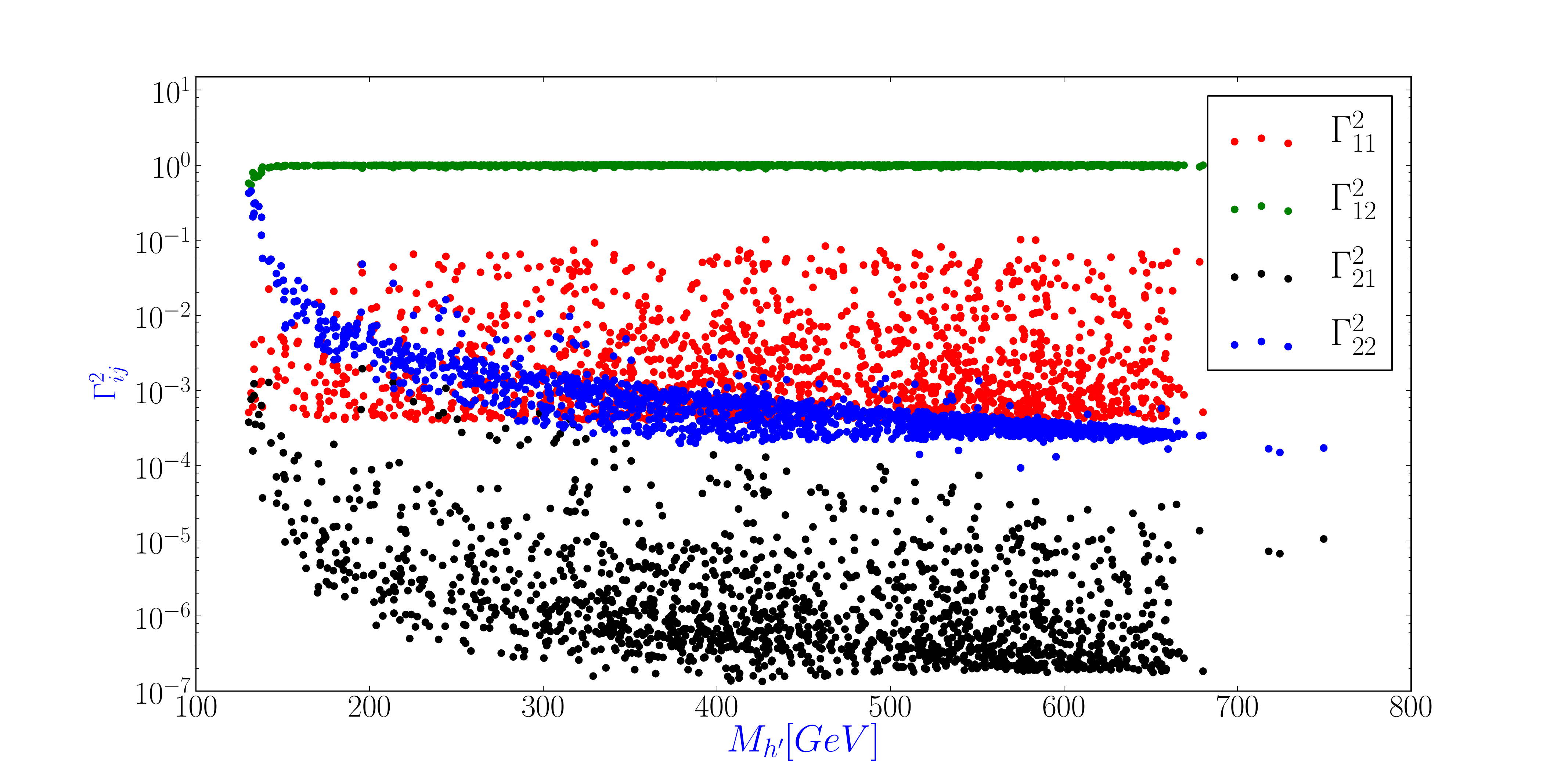}
\caption{Decomposition of the BLSSM Higgs boson, $h'$, and the SM-like Higgs, $h$, versus $M_{h^\prime}$.}
\label{fig:1}
\end{figure}

In Fig.~\ref{fig:1} we show the $h'$ (in $\Gamma_{21}, \Gamma_{22}$) and $h$ (in $\Gamma_{11}, \Gamma_{12}$) decompositions. Note that, if $\tilde{g} = 0 $, the coupling of the BLSSM lightest Higgs boson with the SM particles vanishes at tree level and is very suppressed ($\sim{\cal{O}}(10^{-6})$) at loop level.   Here we choose a parameter space such that the lightest chargino is rather light, $M_{\chi^\pm} = 120 $ GeV, so as to enhance the SUSY contributions to the Higgs decays into $\gamma \gamma$ and $Z \gamma$, namely, we consider a low $\tan \beta$ between $1.1$ and $5$ and $\mu$ and $M_2$ between $100$ and $300$ GeV,  while other  SUSY mass and trilinear parameters are assumed to be of order few TeV. It is worth mentioning that the dominant decomposition for the SM-like Higgs state is $\Gamma_{12} \sim {\cal O}(1)$, which is equivalent to $\sin \beta\sim {\cal O}(1)$ in the MSSM, and that the light BLSSM Higgs, $h'$, is dominated by $\Gamma_{23}$ and $\Gamma_{24} \sim {\cal O}(0.5)$.

\begin{figure}[th]\centering
\includegraphics[height=6.5cm,width=10cm]{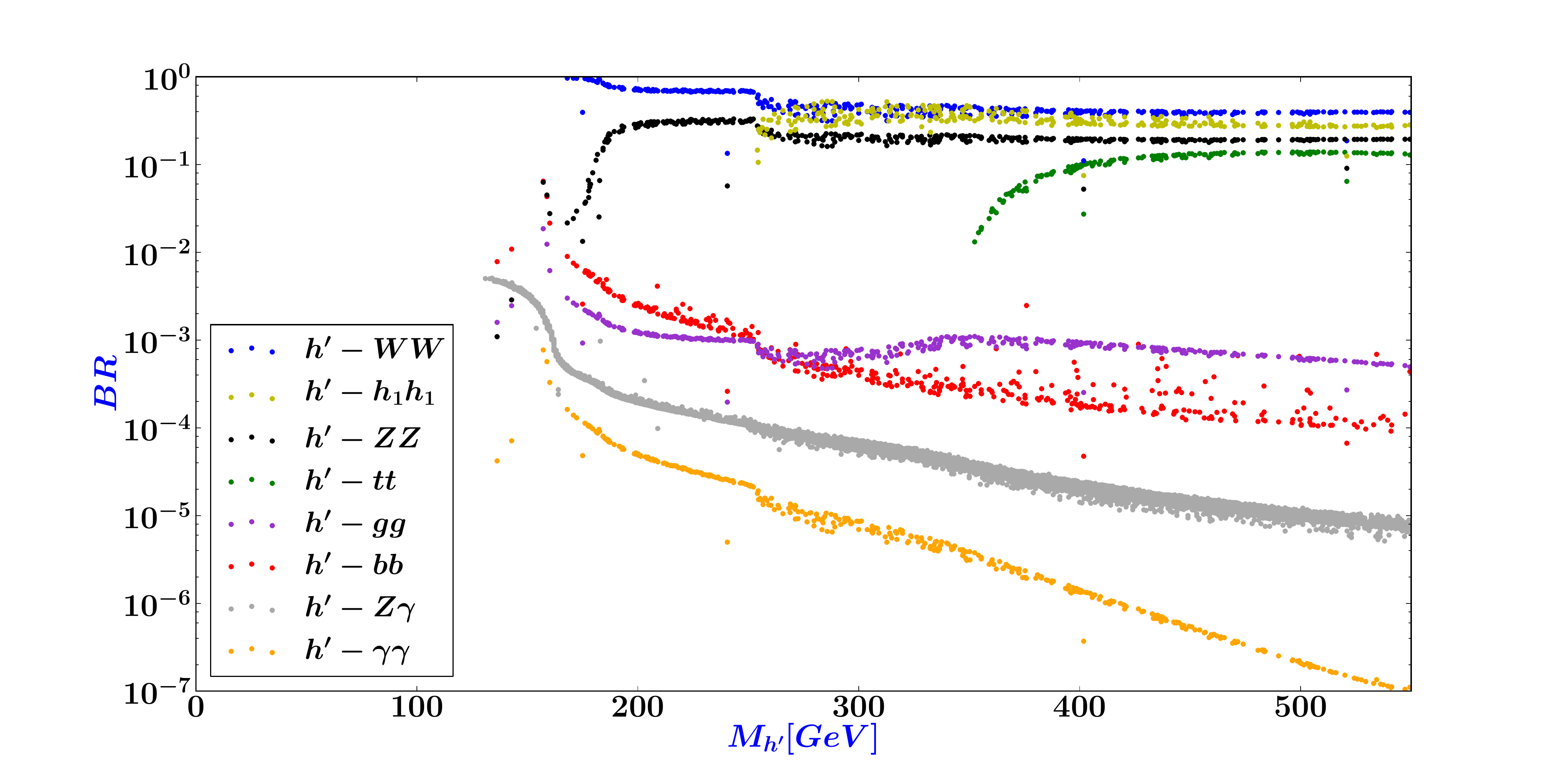}
\caption{The {\rm{BR}}s of $h'$ versus $M_{h'}$ for $0.1\le \tilde{g}\le 0.25$ and $g_{BL} = 0.5$.}
\label{fig:2}
\end{figure}

We display in Fig.~\ref{fig:2} the Branching Ratios ({\rm{BR}}s) of $h^\prime$ into all  its possible decay channels, for non-zero $\tilde{g}$, including $gg$, $\gamma\gamma$ and $Z\gamma$ that are induced at one loop level.  A few remarks on this figure are in order: $(i)$ for $m_{h'} \geq 200$ GeV, $h'$ decays are dominated by the $W^+ W^-$ and $hh$ channels; $(ii)$ in the  BLSSM the {\rm{BR}}$(h' \to Z \gamma)$ is typically larger than  the {\rm{BR}}$(h' \to \gamma \gamma)$, unlike the MSSM and SM where it is the other way around.

\subsection{Implementation and simulation}

 The Higgs production modes included in our forthcoming numerical analysis are gluon-gluon Fusion (ggF), which induce around $90\%$ of the total cross section (hereafter denoted by $\sigma$), while  Vector-Boson Fusion (VBF), Higgs-strahlung (VH) and associated production with top-quarks (ttH) contribute with around $10\%$. 
The data analyses in these channels are based on an integrated luminosity of $20$ fb$^{-1}$ at $\sqrt s=7,8$ TeV and expected to rely upon from 
$100$ to ${\cal O}(1000)$ fb$^{-1}$ at $\sqrt{s}= 13$ TeV. The magnitude of the signal is usually expressed via the ``signal strength'' parameters, {{defined as 
\bea 
\mu_{_{XY}} &=& \frac{\sigma(pp \to h' \to XY)}{ \sigma(pp \to h \to XY)^{\rm SM}} = \frac{\sigma(pp \to h')}{ \sigma(pp \to h )^{\rm SM}}  \times \frac{{\rm BR}(h' \to XY)}{ {\rm BR}( h \to XY)^{\rm SM}}.
\eea}}

\noindent
Herein, the $h'$ in the numerator is indeed the lightest BLSSM CP-even state and the $h$ in the denominator is the SM
Higgs boson with mass 125 GeV whereas in both cases the cross section is intended as computed inclusively (i.e.,
over the ggF, VBF,  VH and ttH modes\footnote{Hereafter, the bulk of the production rates will be due to the first two channels.}). 

For the implementation of the BLSSM we used SARAH \cite{Staub:2013tta} and SPheno \cite{Porod:2011nf} to build the model. For loop induced channels we linked it with CP-SuperH \cite{Lee:2012wa}. The matrix-element calculation and
events generation were derived by MadGraph \cite{Alwall:2011uj}. We then used Pythia \cite{Sjostrand:2007gs} to simulate the initial and final state radiation, fragmentation and hadronisation effects. For detector simulation we passed the Pythia output to Delphes \cite{deFavereau:2013fsa}. For data analysis, we used MadAnalysis5 \cite{Conte:2012fm}. 

In our scans, for the computation of the signal strength distributions  in the next section,
we consider the following regions of the parameter space: 
\bea 
m_0 &=& 1-3~ {\rm TeV}, M_3 = 3~ {\rm TeV},  M_2= 120 - 300~ {\rm GeV},  M_1 = 100 -500~ {\rm GeV},  \tan \beta = 5, \nonumber \\
 \tan \beta' &=&1.15, \vert A_0 \vert = 1.5 - 3~ {\rm TeV}, \mu =100 -350  ~ {\rm GeV}, \vert \tilde{g} \vert = 0.1 - 0.25, g_{BL} =0.5.
\eea
In addition, in the upcoming event generation analyses, the following benchmark point is assumed:
\be \label{benchs}
m_{\chi_1^+} = 120 ~ {\rm GeV}, \mu = 120  ~ {\rm GeV},  \tan \beta =5, \tan \beta' = 1.15 , \tilde{g} =-0.24, g_{BL} =0.5, 
\ee
while all other SUSY particles are of order TeV.  This benchmark point is consistent with current theoretical and experimental limits, as we determined through an independent program checked against specialised literature. {It is worth pointing out  that light $\mu$ and chargino mass are crucial for enhancing the SUSY contributions to $h \to \gamma \gamma$ and $h \to Z \gamma$ simultaneously}. Finally, notice that the $h'$ masses considered below (140, 300, 350 and 480 GeV) are all accessible through the inputs in Eq.~(\ref{benchs}), upon suitable adjustments of the Higgs potential parameters.

\section{Search for a heavy BLSSM Higgs boson at the LHC}
\label{sec:heavy}

In this section we analyse possible signatures of the lightest genuine BLSSM scalar boson $h'$
when it is rather heavy, with mass between 300 GeV and 1 TeV, at Run 2 of the LHC. 
Fig.~\ref{fig:2} shows that the decay channels available to the $h'$ state are the same as those of the SM-like
$h$ one, with the notable exception of the former decaying into (pairs of) the latter, i.e., $h'\to hh$. The corresponding
{\rm{BR}} can be in fact the dominant one, once its threshold is open. It is therefore the distinctive feature of a heavy $h'$ whenever
$m_{h'}\ge 2 m_h$.

 ATLAS \cite{ATLAS:2014rxa} and CMS \cite{CMS:2014eda} have both recently searched for $hh$ signals decaying to a $4b$ final state.  However, it turned out to be a  significant challenge to distinguish the emerging signature, made of 
 of four $b$-jets  in the final state, from the huge multi-jet QCD background. In fact, the sensitivity achieved by the LHC
experiments was rather poor and results obtained were consistent with the SM. We shall nonetheless attempt extracting this signal, so as to compare the scope of  detecting it at Run 2 versus what has been
achieved at Run 1.

The decay $ h' \to h h \to\gamma\gamma b\bar b$, which  has been experimentally analysed in Refs.~\cite{Aad:2014yja,CMS:2014ipa}, may prove to be the best way to probe a heavy $h'$ of the BLSSM, since 
the problem of a suppressed $h\to\gamma\gamma$ decay is offset by the
fact that  both $h'\to hh$ and $h\to b\bar b$ are the dominant decays of the two Higgs states concerned. 
The aforementioned  searches were performed on the $\sqrt{s} = 8$ TeV data set corresponding to an integrated luminosity of $ \approx20~{\rm fb}^{-1}$. Following these, the   
ATLAS collaboration observed five excess events (above and beyond the expected SM yield) 
within a mass windows from  260 to 500 GeV, which represent an excess of $2.4\sigma$, with an intriguing  
$p_0$-value (local probability of compatibility with the background) $\sim 10^{-3}$ at 300 GeV, which corresponds to $3.0~ \sigma$ \cite{Aad:2014yja}. In contrast, CMS reported that searches within the mass region from 260 GeV to 1100 GeV were consistent with expectations from SM processes \cite{CMS:2014ipa}. Needless to say then, we will thoroughly
investigate this signature too at the upcoming Run 2. 

Before proceeding to doing so in two separate subsections, let us start by explaining how such large decay rates for
$h'\to hh$ can occur in the BLSSM. Herein, the scalar trilinear coupling between $h'$ and $h h$ is given by 
\be
\lambda_{h^\prime hh}^{\rm BLSSM} = \frac{-i \tilde{g} g_{_{BL}} }{4} \Gamma_{i2}^2 \left( 2 v'_2 \Gamma_{24} -  v'_1 \Gamma_{23} \right).
\ee
Here we have assumed, as advocated in the previous section, that $\Gamma_{12} \gg \Gamma_{11,13,14}$ and $\Gamma_{23,24} \gg \Gamma_{21,22}$.  This should be compared with the MSSM trilinear scalar coupling 
\be
\lambda_{Hhh}^{\rm MSSM} =  -i \frac{g_1^2+g_2^2}{4} v \left[2 \sin2\alpha \sin(\beta+ \alpha) - \cos 2\alpha \cos(\beta+\alpha)\right],  
\ee 
for which, when $\sin \beta > \cos \beta$ and assuming the decoupling limit where $\alpha \sim \beta$, one finds 
  \begin{equation}\lambda_{Hhh}^{\rm MSSM} =   -i \frac{g_1^2+g_2^2}{4} v \sin^3 \beta . \end{equation}
Also note that the ${Hhh}$ coupling is modified in the BLSSM with respect to the MSSM and takes the form
\be
\lambda_{Hhh}^{\rm BLSSM} = \frac{i }{4} (g_1^2+\tilde{g}^2+ g_2^2) \Gamma_{31} \left( v_d \Gamma_{12}^2 + 2 v_u \Gamma_{12} \Gamma_{11} )\right). 
\ee 
It is clear that $\lambda_{h'hh}^{\rm BLSSM} \propto v'_{1,2} \sim {\cal O}(1)$ TeV  is much larger than the coupling ${Hhh}$ in either SUSY model, which is of order of the EW scale. Therefore, one would expect that the decay rate of $h'$ $\to$ $hh$ is always much larger than that of $H\to hh$. 

\subsection{The $hh\to 4b$ decays of a heavy BLSSM Higgs boson}

The total cross section for the aforementioned $4b$ final state is given by 
\be 
\sigma (pp \to h' \to h h \to 4 b) = \sigma (pp \to h') \times {\rm{BR}}(h' \to hh) \times {\rm{BR}}(h\to b \bar{b})^2,
\ee
and  is dominated by ggF which is in turn obtained as (for a CM energy of 13 TeV)
\begin{equation} \sigma(pp \to h) \times \Gamma_{22}^2 \simeq {\cal O}(1)~ {\rm pb} \end{equation}
while, for $m_{h'} \simeq 350$ GeV, the {\rm{BR}}$(h' \to hh) \sim 0.5$ and the {\rm{BR}}$(h\to b \bar{b})\sim 0.6$, as can be seen from Fig.~\ref{fig:2}.
Thus, one finds that $\sigma (pp \to h' \to h h \to 4 b)$ in the BLSSM  $\sim 10^{-1}~{\rm pb}$. Altough the high total cross section, the huge contribution from background b-jet radiation exceed the signal, so that the associated events would not appear as significant over the SM background. This conclusion is confirmed by  Fig.~\ref{fig:12}, where we show the number of events of signal with its irreducible background  as a function of the invariant mass of the four $b$-jets, $M_{4b}$. 
{{Note that we used the $b$-tagging algorithm included in MadAnalysis \cite{Conte:2012fm}, so that a jet is identified as originating from a $b$-quark when it can be matched to it once it lies within a cone of radius certain $R$ around one of the parton-level $b$-quarks, this yielding an efficiency of about $65\%$.

  Here, we considered the cuts applied in \cite{Aad:2015uka}: i.e.,
{candidate events are required to have at least four $b$-tagged jets, each with $p_T \ge 40$ GeV and separated by a cone of $\Delta R = 1.5$.}
However, as can be seen from the plot,  the signal is well below the background. The highest background contribution {comes from a muti-jets final state, followed by $t\bar{t}$ production and  (semi-)hadronic (anti)top decays which gives about $22\%$ of the noise while the reducible background contributions come from `$Z$ + jets', $ZZ$ and $Zh$ and are found to contribute  less than $1\%$.  }

The signal distribution is presented for $m_h'\approx 2 m_t\approx 350$ GeV, which is in fact the worse case scenario,
as this is where the $t\bar t$ background peaks in $M_{4b}$. However, we have tried different $m_{h'}$ values, to 
no avail, in the mass range from 300 GeV to 1 TeV. The signal would never be accessible, neither with standard nor
with upgraded luminosities. 

\begin{figure}[!h]\centering
\includegraphics[width=8cm,height=6cm,angle=0]{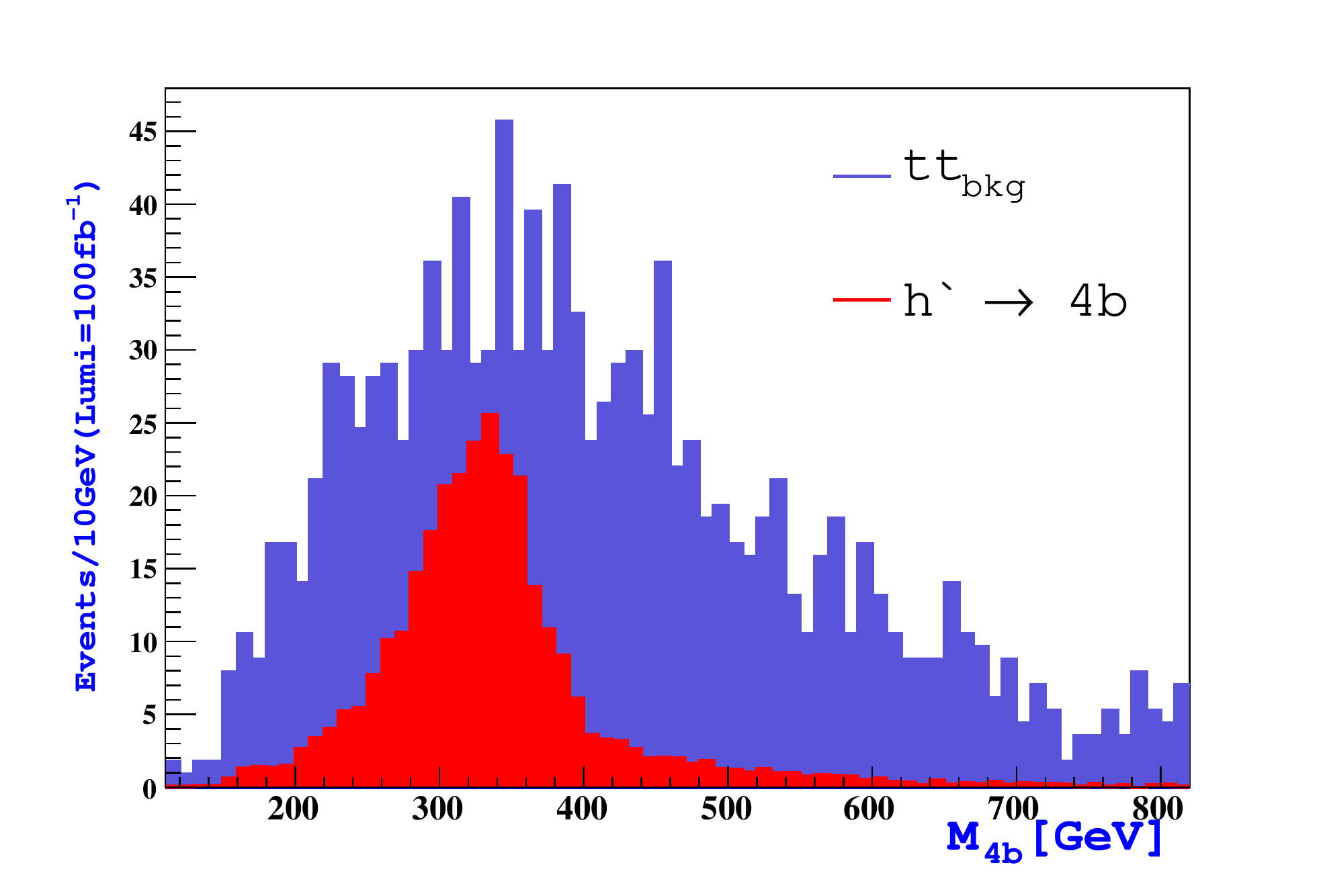}
\caption{Number of signal events for $h^\prime \to hh\to 4b$ decays (red) 
induced by ggF and VBF versus the $4b$ invariant mass
at $\sqrt s=13$ TeV after 100 fb$^{-1}$ of luminosity 
alongside the {$t \bar t$ background (blue).
(The huge multi-jet background, which is given in Ref. \cite{Aad:2015uka}, is not shown.)} 
Here, $m_{h'}=350$ GeV.}
\label{fig:12}
\end{figure}

\subsection{The $hh\to b\bar{b}\gamma\gamma$ decays of a heavy BLSSM Higgs boson}

Now we turn to the process $pp \to h' \to hh\to\gamma\gamma b\bar b$. Although this mode  has smaller cross section than $\sigma (pp \to h' \to hh\to 4 b)$, it is more promising due to the clean di-photons trigger with excellent mass resolution and  low background contamination. This is confirmed in Fig.~\ref{fig:13}, where the number of signal events is displayed versus the background as a function of the invariant mass $M_{\gamma \gamma bb}$ for two examples of $h'$ masses: $m_h'=300$ GeV and $m_{h'} = 480$ GeV.  

\begin{figure}[ht]\centering
\includegraphics[width=8cm,height=6cm,angle=0]{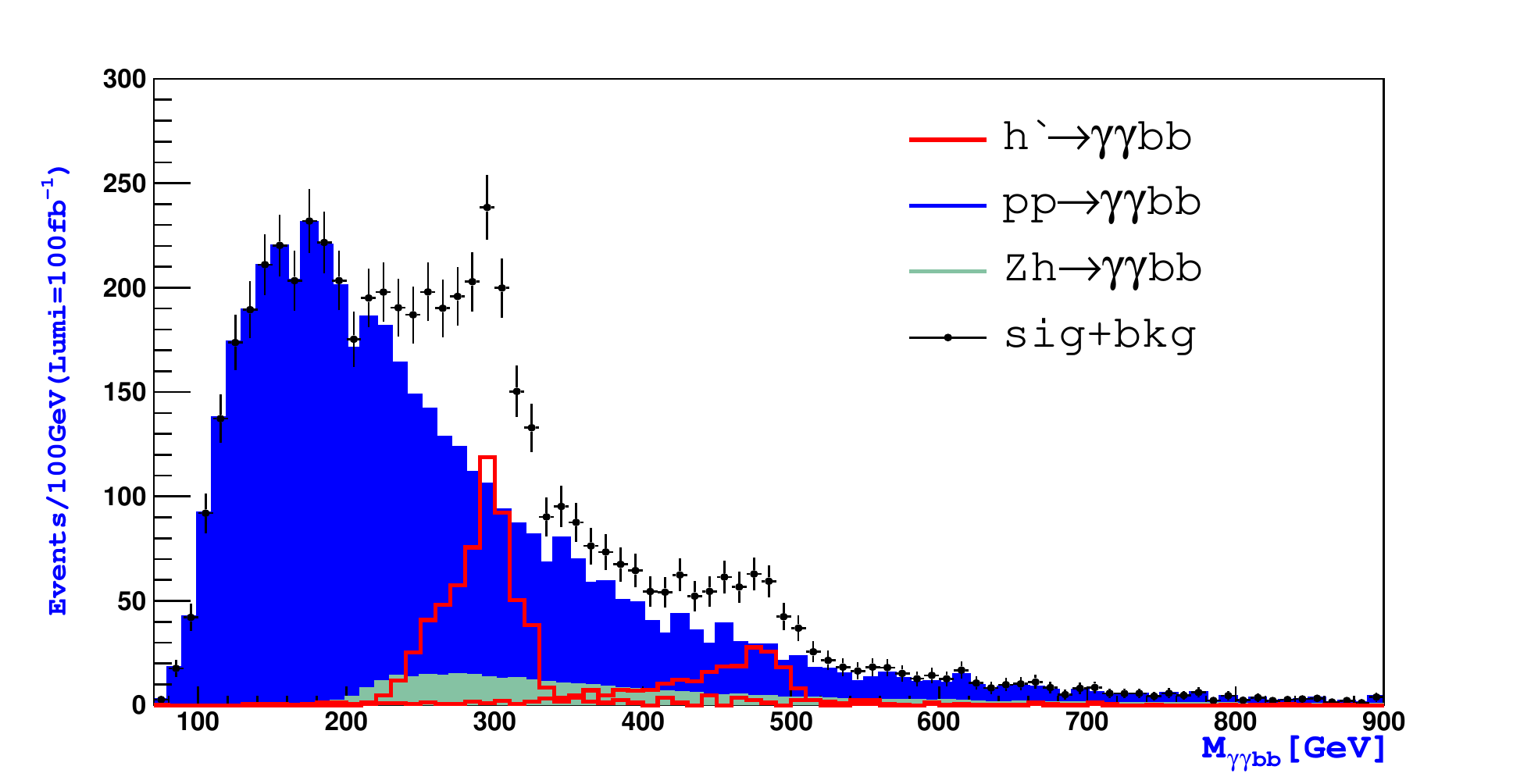}
\caption{Number of signal events for $h^\prime \to\gamma\gamma b\bar b$ decays (red) 
induced by ggF and VBF versus the $\gamma\gamma b\bar b$ invariant mass
at $\sqrt s=13$ TeV after 100 fb$^{-1}$ of luminosity 
alongside the two dominant $\gamma\gamma$ (blue) and $Zh$ (green) backgrounds. Their sum is also shown as data points. Here, $m_h'=300$ and $480$ GeV.
}
\label{fig:13}
\end{figure}

The background to this process can be classified  into two categories: background events containing a real Higgs boson decay, $h\to\gamma\gamma$ and $h\to bb$, and the continuum background of events not containing a Higgs boson. The continuum contribution in the signal region is split between events with two photons and events with a single photon in association with a jet faking the second photon. Further, the two $b$-tagged jets include real heavy-flavour jets as well as mis-tagged light-flavour jets and gluons. The contribution from di-leptonic decays of $t\bar{t}$ events where two electrons fake the two photons is roughly 10$\%$ of the total background. {{The contribution from other processes, like leptonic decays of di-gauge bosons  where two electrons fake the two photons and the Higgs boson comes associated with a $W/Z$, is negligible}}. In our analysis, we adopt the following acceptance cuts in transverse momentum, pseudorapidity of and separation amongst the photons and jets:
{\begin{enumerate}
\item the pseudorapidity $\eta$ of the two photons must fall within the geometric acceptance of the detector for photons, $|\eta | \le 2.4$;
\item the ratio between the transverse momenta of the leading and subleading photon must be $\ge 0.25$;
\item jets are required to fall within the tracker acceptance of $|\eta| \le 2.5$ with transverse momentum $p_T \ge 35 $ GeV.

\end{enumerate}}}
After our preselection is enforced, 
already at standard luminosity of Run 2, the signal is clearly visible above all backgrounds, both at 300 and 480
GeV,  thereby enabling one to declare discovery of a Higgs-to-two-Higgs signal as well as circumstantial evidence of
a BLSSM decay chain of the type $h'\to hh$.  In order to eventually profile the latter though, the simultaneous reconstruction of the two $h$ resonances and of the $h'$ one is a pre-requisite. To this end, in 
 Fig.~\ref{fig:hh_h}, we also show the mass reconstruction of the two SM-like Higgs boson masses, in the two
channels $h\to\gamma\gamma$ and $b\bar b$, against the backdrop of the SM noise. From the corresponding
distributions, a clear element emerges that characterises this signature is very promising, i.e., the very efficient
reconstriction of $m_h\approx125$ GeV from the di-photon pair, from which is evident the strong background suppression
which can be achieved. In contrast, this is not true in the case of $b\bar b$ decays, as here  the background remains overwhelming above the signal (implicitly also explaining the reduced sensitivity of the fully hadronic $4b$ signal 
previously considered, where jet combinatorics would further play a significant role in degrading the quality of it).
Furthermore, notice that the quality of the mass reconstruction is not dramatically different for $m_{h'}=300$ and $480$ GeV.

\begin{figure}[!h]\centering
\includegraphics[width=7.5cm,height=6cm,angle=0]{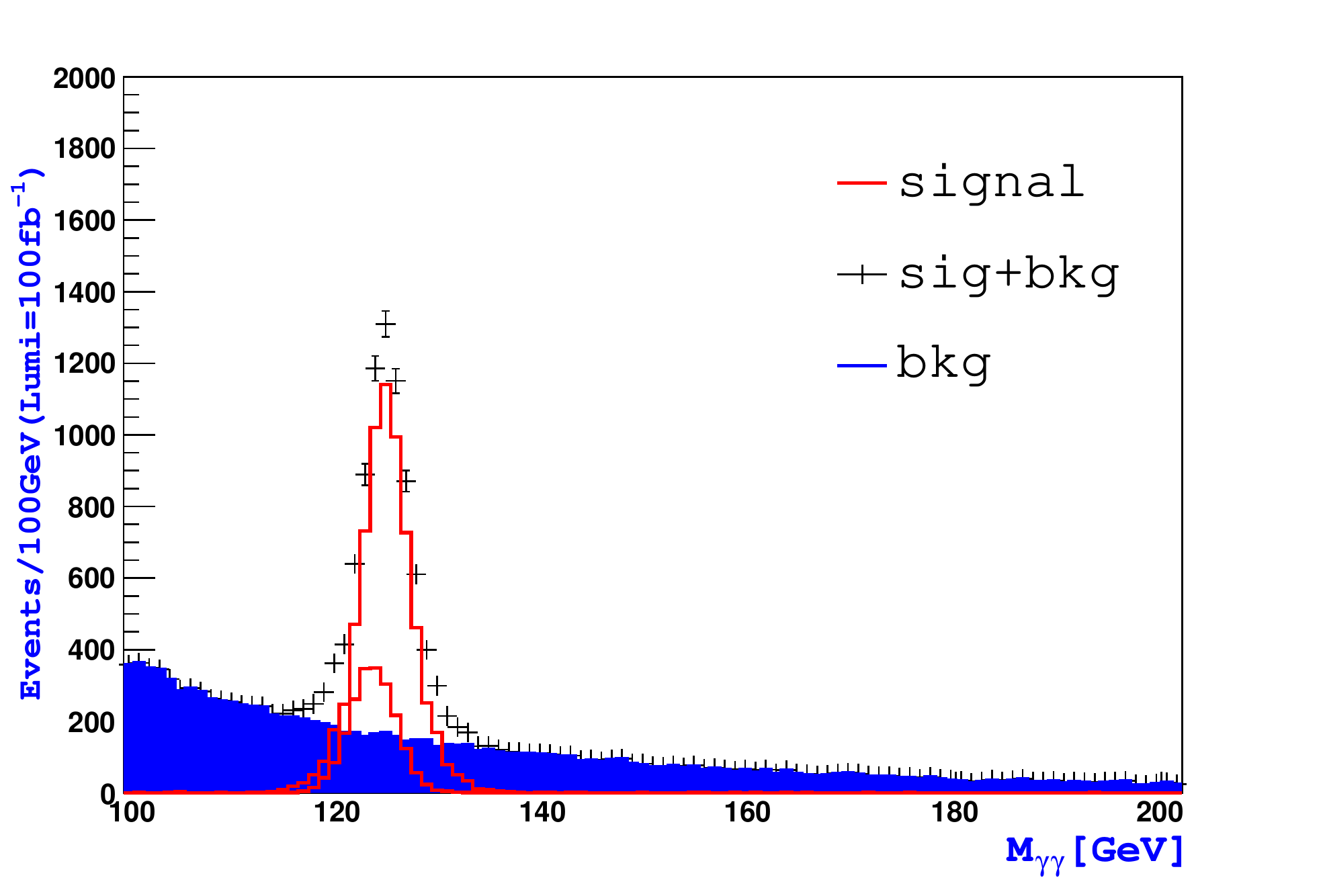}~~\includegraphics[height=6cm,width=7.5cm,angle=0]{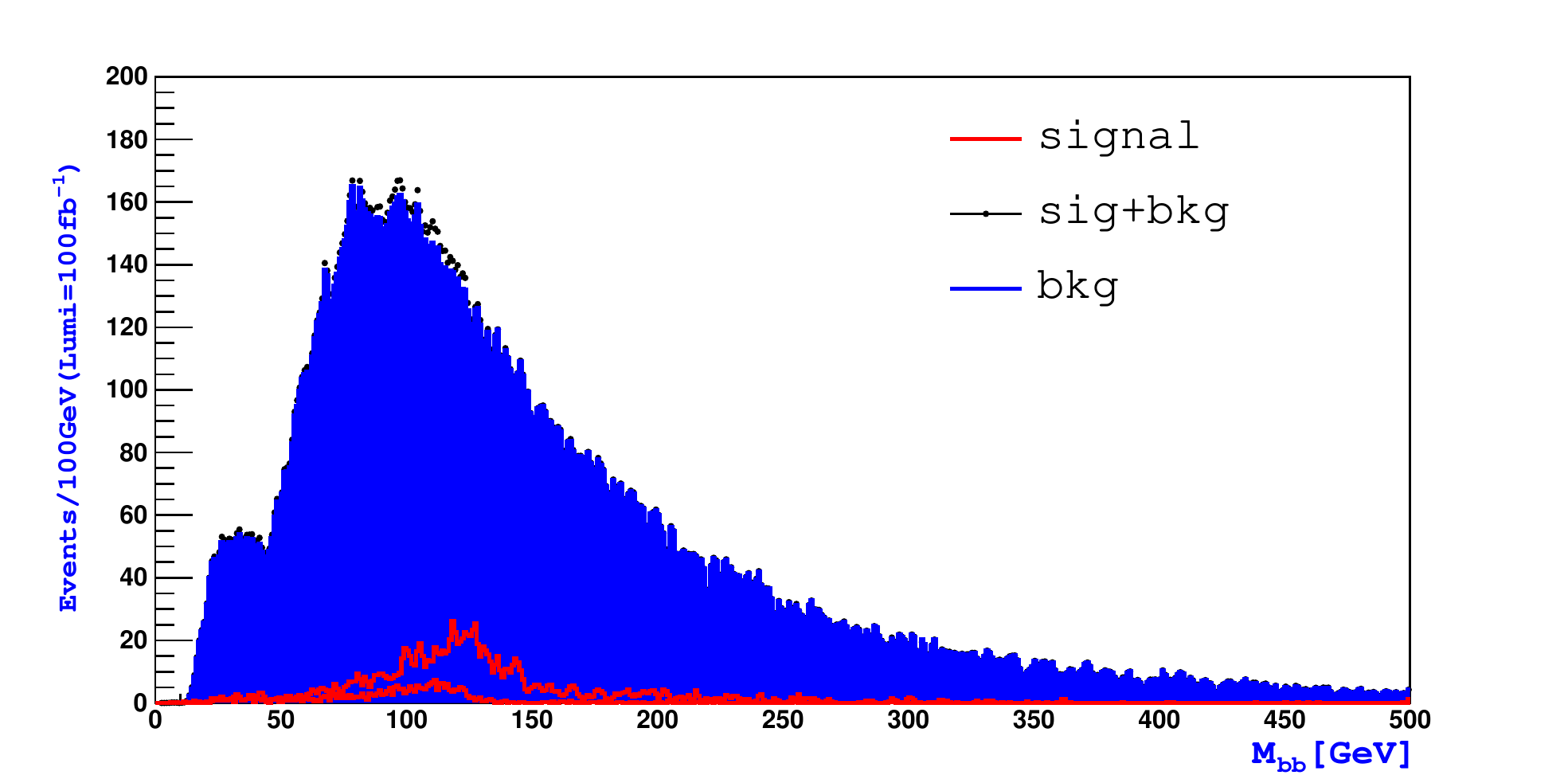}
\caption{Number of signal events for $h^\prime \to\gamma\gamma b\bar b$ decays (red) 
induced by ggF and VBF versus the $\gamma\gamma$ (left) and $ b\bar b$ (right)  invariant mass
at $\sqrt s=13$ TeV after 100 fb$^{-1}$ of luminosity 
alongside the total background (blue). Their sum is also shown as data points. Here, $m_h'=300$ and $480$ GeV.
Only the acceptance cuts described in the text are used here.}
\label{fig:hh_h}
\end{figure}

In the light of the mass distributions just discussed, one can attempt a more refined signal selection against the
continuum noise. In Tab. 2  we show the number of events for signal and continuum background after each cut mentioned
therein  and Fig.~\ref{fig:1_cut} shows the final number of events versus the background after all cuts are applied. It is clear from this plot that the final result is an almost background-free $M_{\gamma\gamma b\bar b}$
distribution neatly pointing to the value of the $h'$ mass, for values between 300 and 480 GeV. 
It is not surprising then, in the end, significances for the signal can be extremely large, as seen in Fig.~\ref{fig:lumi2b2a},
for any $m_{h'}$ value, after a final 
sampling in $M_{\gamma\gamma b\bar b}$
is exploited. Notice that, here, both reducible and irreducible backgrounds are accounted for in the calculation.

\begin{table}[!h]
\centering
  \label{table-bbAA}
  \begin{center}
        \begin{tabular}{c|c|c|c}
     Applied cut & Signal, $m_{h^\prime}= 300$ & Signal,  $m_{h^\prime}= 480$ & Continuum background\\
      \hline
 {After acceptance cuts}  & 626 & 237 & 4758 \\
        \hline
  $M_{\gamma\gamma} \le 135$ GeV & 625 & 234 & 4375  \\
\hline
  $M_{\gamma\gamma} \ge 115$ GeV & 616 & 223 & 182  \\
  \hline
 $M_{b\bar b} \le 145$ GeV & 536 & 210 & 98  \\
  \hline
  $M_{b\bar b} \ge 105$ GeV & 351 & 86 & 30  \\
  \hline
  \end{tabular}
\caption{Signal (for two $h'$ mass values) and continuum background events in the $\gamma\gamma b\bar b$ channel 
as a function of  several mass selection cuts.
The energy is $\sqrt s=13$ TeV whereas the luminosity is 100 fb$^{-1}$.}
  \end{center}
  \end{table}

\begin{figure}[h!]\centering
\includegraphics[width=8cm,height=6cm,angle=0]{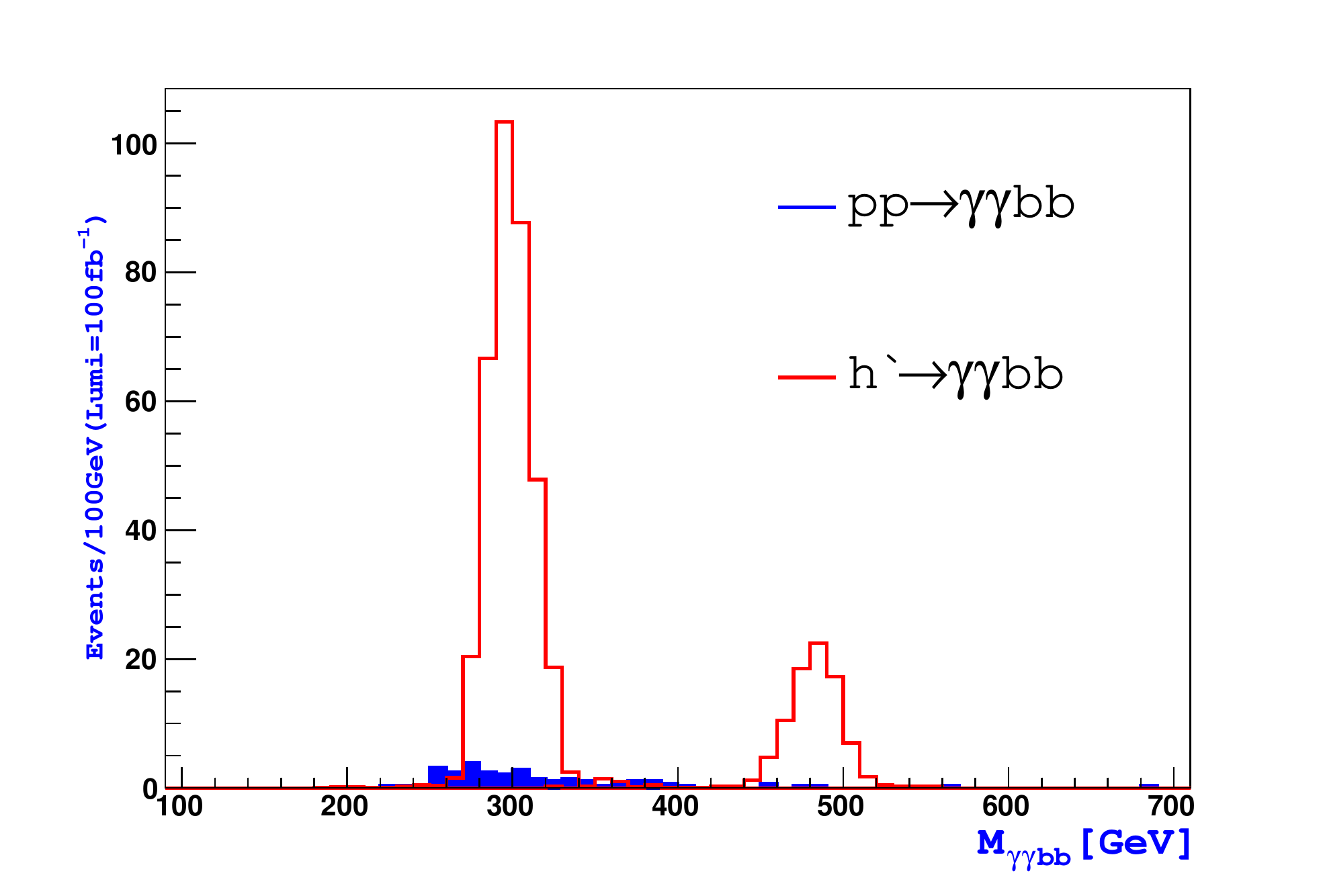}
\caption{Number of signal events for $h^\prime \to\gamma\gamma b\bar b$ decays (red) 
induced by ggF and VBF versus the $\gamma\gamma$ (left) and $ b\bar b$ (right)  invariant mass
at $\sqrt s=13$ TeV after 100 fb$^{-1}$ of luminosity 
alongside the total background (blue). Here, $m_h'=300$ and $480$ GeV.
Also the selection cuts of Tab.~2 are used here.}\label{fig:1_cut}
\end{figure}

\begin{figure}[!h]\centering
\includegraphics[height=6.0cm,width=16cm]{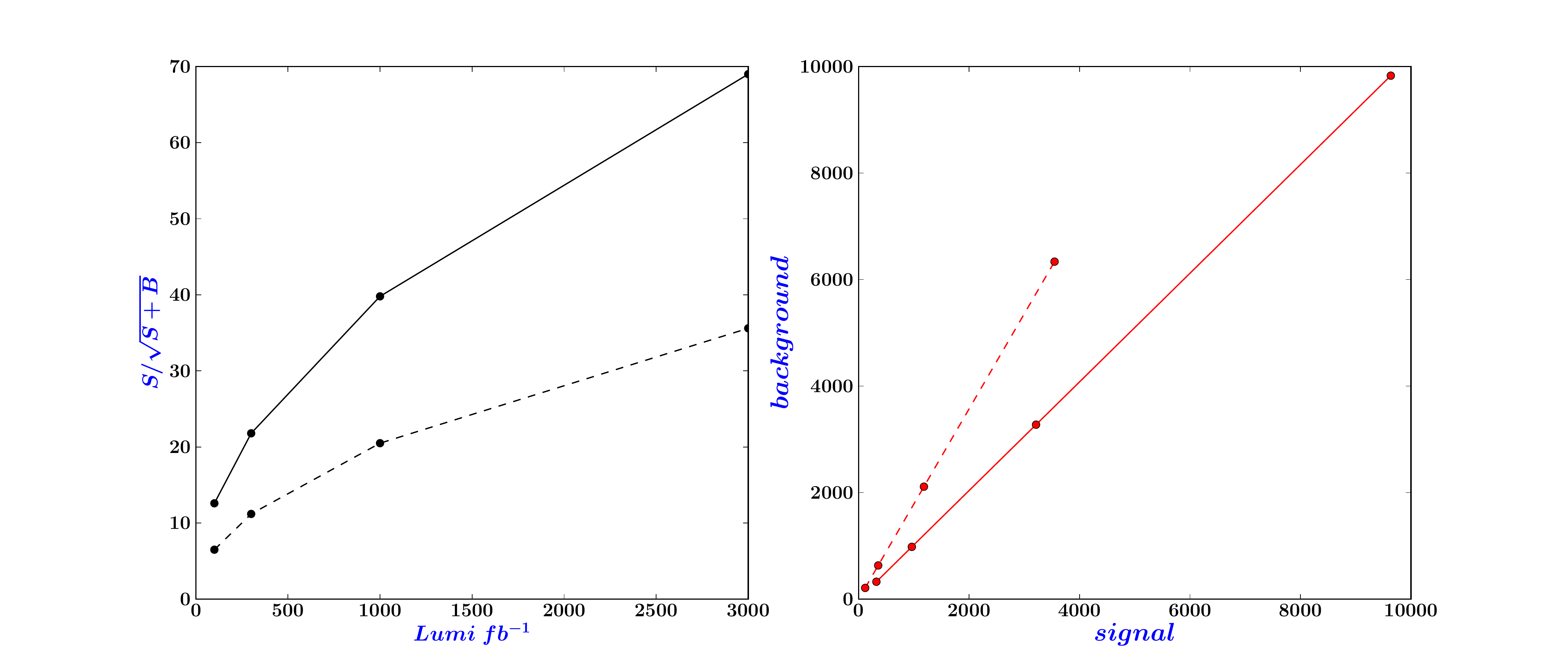}
\caption{Left: Significance of the $h^\prime \to \gamma\gamma b\bar b$ signal (for $m_{h'}=300 $ and 480 GeV)
versus the luminosity (black). Right: Number of events for signal and background for variable luminosity (red). 
Data are produced 
at $\sqrt{s} = 13$ TeV and the  points correspond to an integrated luminosity of 100, 300, 1000 and 3000 fb$^{-1}$.
Notice that event rates are computed after the acceptance cuts described in the text and the mass selections of Tab.~2.
The $M_{\gamma\gamma b\bar b}$ mass windows used for the calculation is 50 GeV for $m_{h^\prime} = 300 $ GeV and 100 GeV for $m_{h^\prime} = 480$ GeV.}\label{fig:lumi2b2a}

\end{figure}

\section{Search for a light BLSSM Higgs boson at the LHC}
\label{sec:light}

In this section we briefly revisit the possible signatures of a light BLSSM Higgs boson $h'$ (with mass $m_{h'} \approx 140$ GeV) at the LHC. As emphasised in Refs. \cite{Abdallah:2014fra,Hammad:2015eca,Khalil:2015vpa}, this particle can be probed in one of the following channels: $\gamma\gamma$, $Z \gamma$ and $ZZ$.  We review these in the three upcoming subsections.

\subsection{The $\gamma\gamma$ decays of a  light BLSSM Higgs boson}

The coupling of a Higgs boson with di-photons  is induced by loops of charged particles. In the SM, these loops are mediated by the $W$ gauge boson and top-quark. In SUSY models, the $h\gamma\gamma$ triangle coupling contains additional loops of charged particles: charged Higgses $H^\pm$ , squarks $\tilde{q}$, sleptons $ \tilde{\ell}^\pm$ and charginos~ $\chi^\pm$. Since  the Higgs boson coupling with SUSY particles are not proportional to their masses  their contributions decouple for high masses. In this paper, we focus on the cases of light charginos, $\chi^\pm_1$, enhancements, since they can increase the $h\gamma\gamma$ amplitude squared up to $30\%$ \cite{Belyaev:2013rza,Hemeda:2013hha} (i.e.,
 {{the sfermions and charged Higgs bosons are assumed to be heavy}}). 

The Higgs decay into di-photons provides a clean final-state topology which allows for the mass to be
reconstructed with high precision. The  partial decay width of the lightest BLSSM Higgs boson into di-photons is given by 
\begin{equation}
  \Gamma(h^\prime \to \gamma\gamma) = \frac{G_\mu \alpha^2 m^3_{h^\prime}}{128\sqrt{2}\pi^3}\left |{{A}}_t + {{A}_W}+ A_{H^\pm} + A_{\tilde f} +{{A}_{\chi^\pm}}\right |^2 , 
\end{equation}
where the amplitudes $A_{f, W, H^\pm, {\tilde f}, \chi^\pm}$ can be found in \cite{Djouadi:2005gj}.  In Fig.~\ref{fig:4} we show  the signal strength of $gg\to h^\prime \to \gamma \gamma$ for 110 GeV $<m_{h'}<$  $150$ GeV. We also include the di-photon signal strengths of the SM-like Higgs, $h$,  in the MSSM and BLSSM, in addition to the MSSM-like heavy Higgs, $H$. It is interesting to note that the BLSSM results for both $h$ and $h'$ are matching  the observed data at Run 1, whereas the signal strength of the heavy Higgs in the MSSM, $H$, is quite suppressed and cannot easily account for these observations. 

\begin{figure}[!h]\centering
\includegraphics[height=7cm,width=10cm,angle=0]{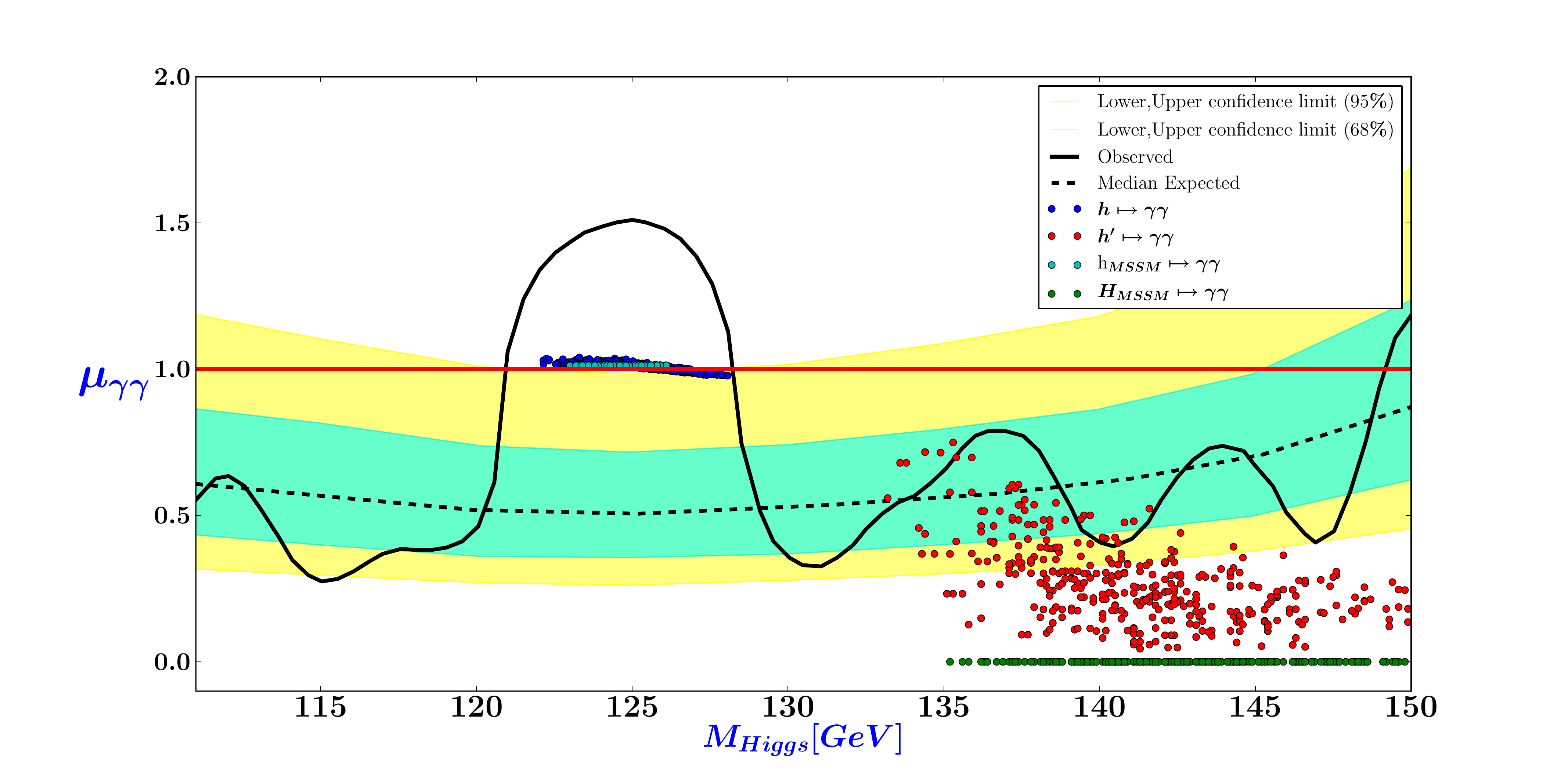}
\caption{Signal strength of the lightest and next-to-lightest Higgs bosons in the BLSSM (in blue and red, respectively) in the $\gamma\gamma$ channel. For comparison, we also include the signal strength of the lightest and next-to-lightest Higgs bosons in the MSSM (in cyan and black, respectively).
The $1$ and $2 \sigma$ confidence intervals are extracted from data collected during 
Run 1 with the observed exclusion limit as given in \cite{Aad:2014eha} is also included. }\label{fig:4}
\end{figure}

The number of events for $h' \to \gamma \gamma$ as function of the  di-photon invariant mass is presented in Fig.~\ref{fig:5}, for a Center-of-Mass (CM) energy $\sqrt{s} = 13\ {\rm TeV}$ and integrated luminosity = $100~{\rm fb}^{-1}$.  Here we  choose the input parameters such that the SM-like Higgs boson has a mass $m_{h}  = 125 \ {\rm GeV}$ and the lightest genuinely BLSSM Higgs state has a mass $m_{h^\prime} \sim 140\ {\rm GeV}$. The dominant backgrounds  consist of an irreducible fraction from  prompt di-photon production and a reducible one from $\gamma +$ jet and di-jet events where one or more of the objects reconstructed as a photon corresponds to a jet, according to CMS ``fake rates''. It is also worth mentioning that here we  consider all cuts applied in the CMS analysis of Ref. \cite{Aad:2014eha}: i.e,
{the photon candidates are collected within  $|\eta^\gamma | \le 2.5$ with  transverse momentum $p_T ^\gamma \ge 20$ GeV.} The production is considered here as induced from both ggF and VBF (as at higher energies the latter
mode grows in importace relatively to the former) and yield both a 
$h$ and $ h^\prime$ state. As can be seen from this figure, the peak at $\sim 140$ GeV is greatly overwhelmed by the background after 100 fb$^{-1}$, yet accessible with additional luminosity, as shown in Fig.~\ref{fig:6}.

\begin{figure}[!h]\centering
\includegraphics[height=6cm,width=8cm]{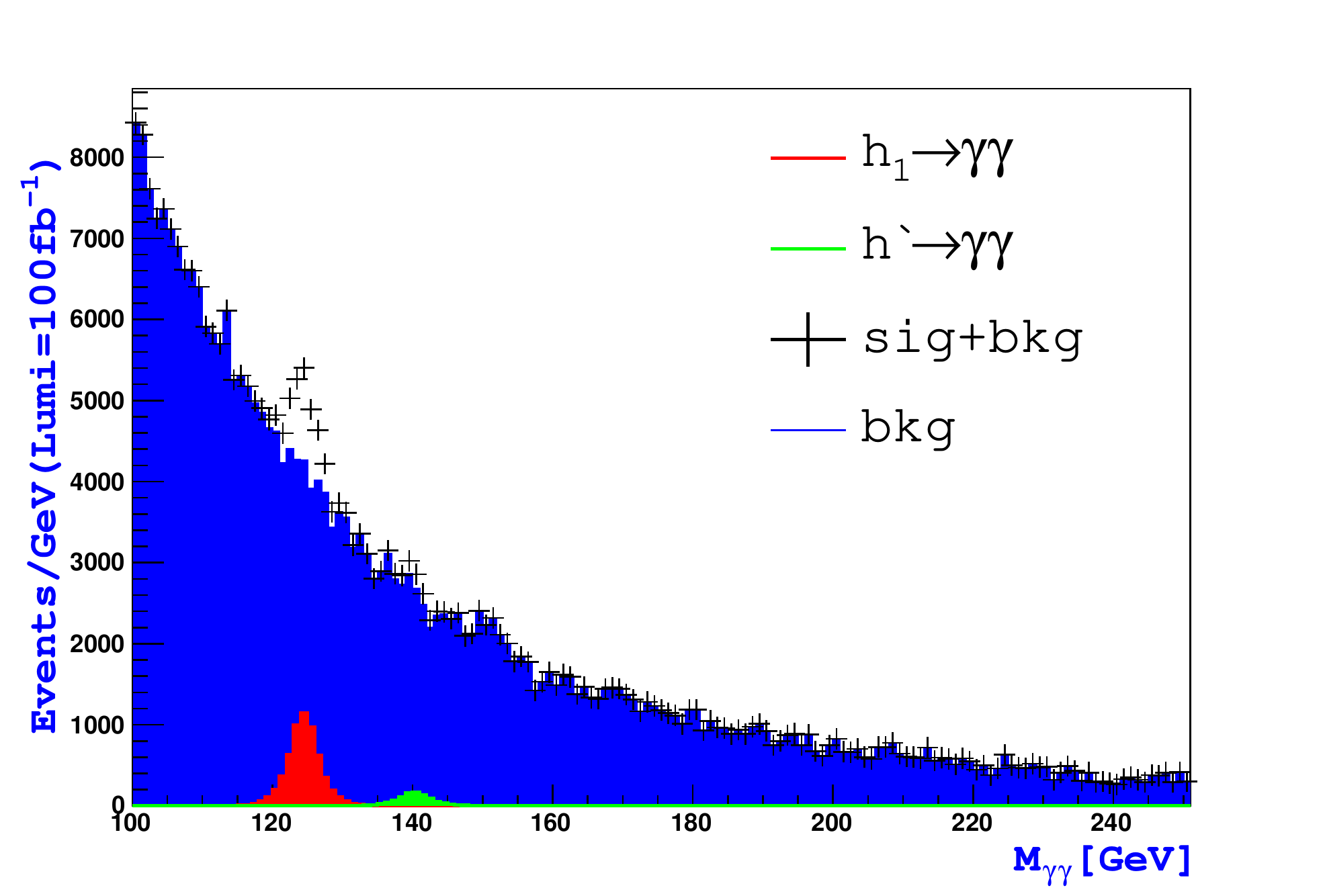}
\caption{Number of signal events for $h$ and $h^\prime \to\gamma\gamma$ decays (red and green, respectively) 
induced by ggF and VBF versus the $\gamma\gamma$ invariant mass
at $\sqrt s=13$ TeV after 100 fb$^{-1}$ of luminosity 
alongside the total background (blue). Their sum is also shown as data points.}\label{fig:5}
\end{figure}

\begin{figure}[!h]\centering
\includegraphics[height=6cm,width=16cm]{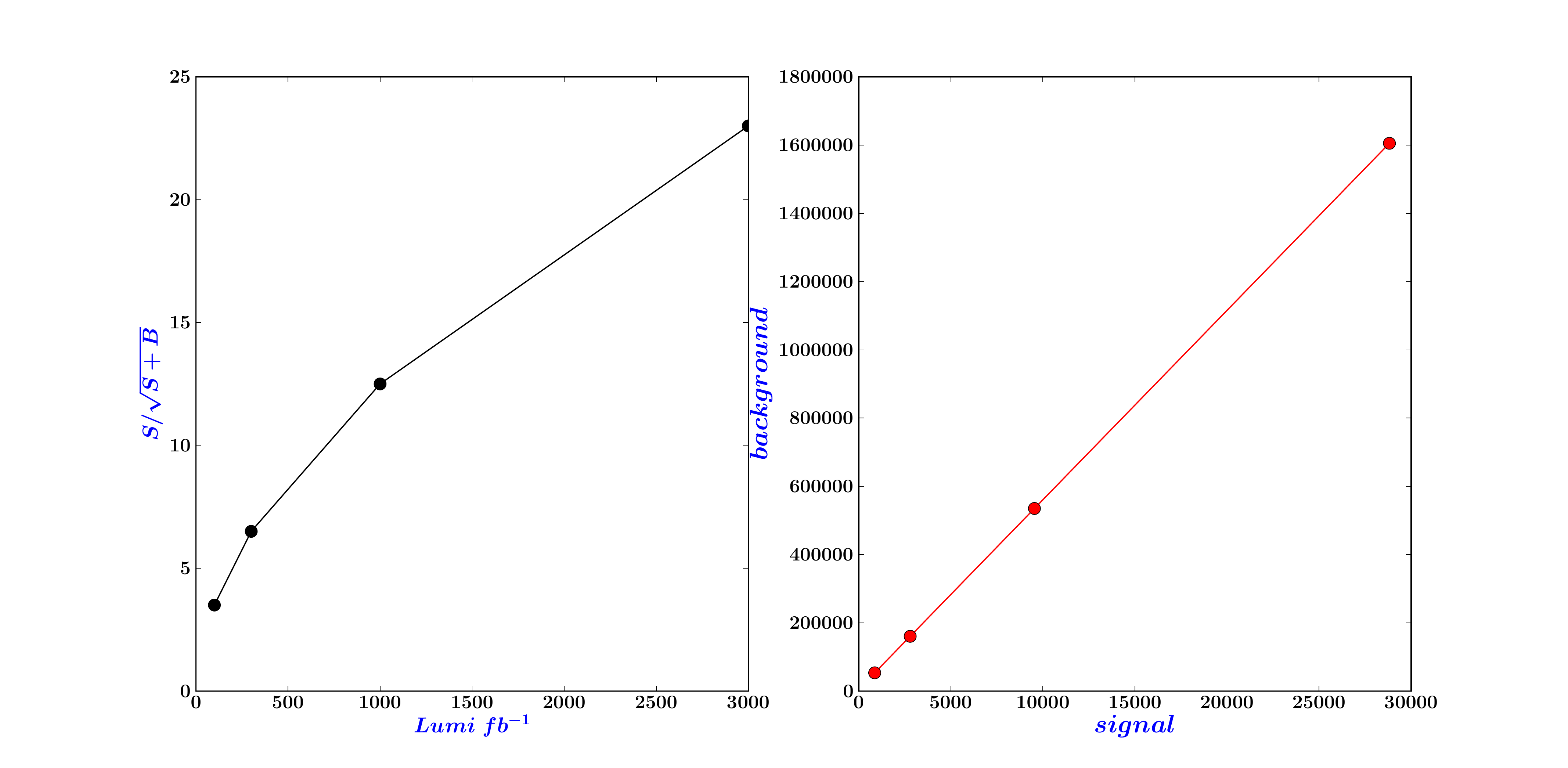}
\caption{Left: Significance of the $h^\prime \to\gamma\gamma$ signal (for $m_{h'}=140 $ GeV)
versus the luminosity (black). Right: Number of events for signal and background for variable luminosity (red). 
Data are produced 
at $\sqrt{s} = 13$ TeV and the  points correspond to an integrated luminosity of 100, 300, 1000 and 3000 fb$^{-1}$.
Notice that event rates are computed after the cut $|m_{\gamma\gamma}-m_{h'}|<10$ GeV.}\label{fig:6}
\end{figure}

\subsection{The $Z(\to \ell^+\ell^-) \gamma$ decays of a light BLSSM Higgs boson}
Despite its small {\rm{BR}}, the LHC experiments 
are currently sensitive to this channel and will be so more and more as luminosity accrues. Precisely because the 
SM rate in this decay channel is small, ATLAS and CMS may access BSM physics through it, owing to the fact that 
the partial width can increase sizeably in presence of additional  loops of charged particles, just like in the $h'\to\gamma\gamma$ channel. The partial decay width of the lightest BLSSM Higgs boson into $Z\gamma$ is given by 
\begin{equation}\label{eq:1bis}
\Gamma(h^\prime\to Z\gamma) = \frac{G^2_f \alpha^2 M^2_W M^3_{h'}}{64\pi^4}\left(1-\frac{M^2_Z}{M^2_{h'}}\right)^3 \left|A_f + A_W + A_{H^\pm} + A_{\tilde f} + A_{\chi^\pm}\right|^2 , \end{equation}
where the amplitudes $A_{f, W,  H^\pm, {\tilde f}, \chi^\pm}$ can be found in \cite{Djouadi:1996yq}. As discussed in \cite{Hammad:2015eca}, due to the mixing in the sfermion and chargino sectors, the diagonal coupling only enhances the $h^\prime \to \gamma\gamma$ channel, while the fact that the $Z$ boson has both vector and axial vector quantum numbers makes both diagonal and off-diagonal couplings of sfermions and charginos contribute to the $h^\prime \to Z\gamma$ channel.  As in $h^\prime \to \gamma\gamma$, we focus here on a light chargino in order to  enhance the $h^\prime \to Z\gamma$ amplitude.

In Fig.~\ref{fig:7} we show that the signal strength of the $h'$, $h$ (both in the MSSM and BLSSM) and $H$ decays to $Z \gamma$ for $m_{h',H}$ around $140$ GeV (as usual, $m_h=125$ GeV), with the 1 and $2\sigma$ confidence intervals extracted from data collected during Run 1 and with the observed exclusion limit as given in \cite{Aad:2014fia}. As can be seen, again,  the BLSSM results for both $h$ and $h'$ match with the observed data rather well whereas the signal strength of the heavy Higgs in the MSSM, $H$, as expected, is quite suppressed, hence unable to reach out to current experimental
results.

\begin{figure}[!h]\centering
\includegraphics[height=7cm,width=10cm,angle=0]{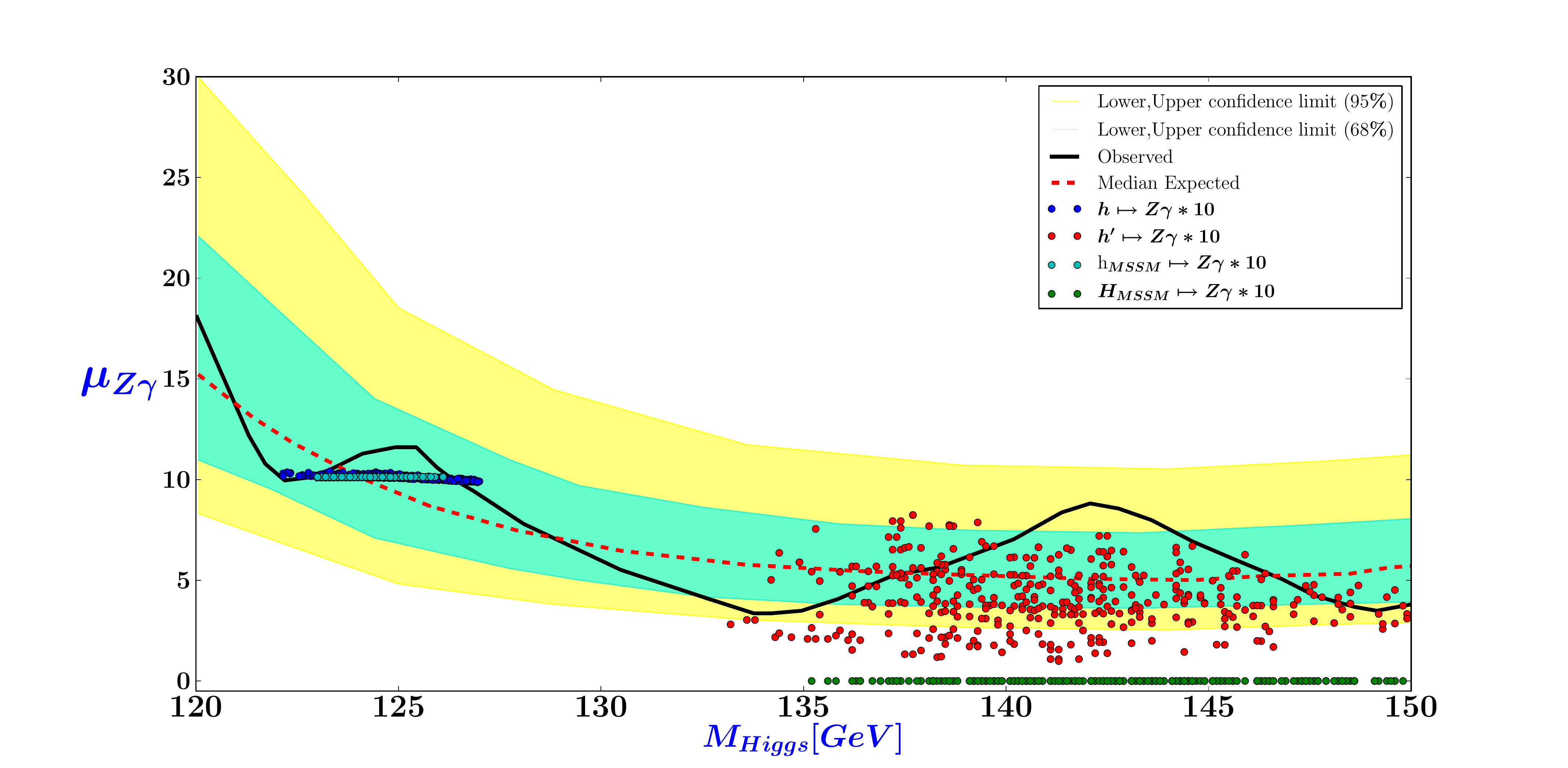}
\caption{Signal strength of the lightest and next-to-lightest Higgs bosons BLSSM (in blue and red, respectively) in the $Z\gamma$ channel. The signal strength of  lightest and next-to-lightest Higgs bosons in the MSSM are given in cyan and black points, respectively.
The signal strength of the lightest and next-to-lightest Higgs bosons in the MSSM are given in cyan and black, respectively.
The $1$ and $2 \sigma$ confidence intervals are extracted from data collected during 
Run 1 with the observed exclusion limit as given in \cite{Aad:2014fia}  is also included.}\label{fig:7}
\end{figure}

The distribution of the `di-lepton + photon' (we assume $Z\to \ell^+\ell^-$, $\ell=e,\mu$)
invariant mass is presented in  Fig.~\ref{fig:8} for the signal and background, where the dominant components of the
latter consist of the irreducible contribution from  $Z\gamma$ production, the reducible one from final state radiation in the neutral Drell-Yan process and `$Z$ + jets' processes where a jet is misidentified as a photon. Here the cuts applied are as in Ref. \cite{Aad:2014fia}, i.e.: 
\begin{enumerate}
\item the photon pseudorapidity must be $|\eta^\gamma | \le 2.5$;
\item the photon transverse momentum must be $ p_T^{\gamma}\ge \ 25 $ GeV; 
\item the di-lepton invariant mass must be $ 85$ GeV $ \le M_{\ell^+\ell^-} \le 95 $ GeV;
\item the `di-lepton + photon'  invariant mass must be $ 130$ GeV $ \le M_{\ell^+\ell^- \gamma} \le 150 $ GeV.
\end{enumerate}
The cut flow results are found in Tab.~\ref{tab:ZA}. The selection (based on Run 1 cuts) remains effective at Run 2 as well,
since already at standard luminosity there could already be an evidence of the $h'\to Z\gamma$ signal in the BLSSM.
The line-shape of the signal, initially swamped by the background (see left-hand side of Fig.~\ref{fig:8}),  would also be very distinctive after the selection is enforced   (see right-hand side of Fig.~\ref{fig:8}). As the luminosity at Run 2 accumulates, the evidence will eventually turn into clear discovery (see Fig.~\ref{fig:ZA}).

\begin{table}[!ht]\centering
    \label{tab:table1} \resizebox{0.75\textwidth}{!}{
    \begin{tabular}{c|c|c|c}

      & Signal ($S$) & Background ($B$) & $\frac{S}{\sqrt{S+B}}$\\
      \hline
     Before cuts  & 200 & 18828 & 1.44 \\
	\hline
  $p_T^\gamma \ge 25$ GeV     & 180 & 6490 & 2.2  \\
\hline
$ 85$ GeV $ \le M_{\ell^+\ell^-} \le 95$ GeV   &172 & 4500 & 2.5  \\
\hline
{$ 130$ GeV $ \le M_{\ell^+\ell^- \gamma} \le 150 $ GeV}  &170 & 3822  & 2.7  \\
  \hline
    \end{tabular}}
\caption{Signal and background events in the $Z\gamma$ channel assuming electron and muon decays of the $Z$ boson
as a function of the selection cuts detailed in the text. The energy is $\sqrt s=13$ TeV whereas the luminosity is 100 fb$^{-1}$.}
 \label{tab:ZA}
\end{table}

\begin{figure}[!h]\centering
\includegraphics[width=7.5cm,height=6cm,angle=0]{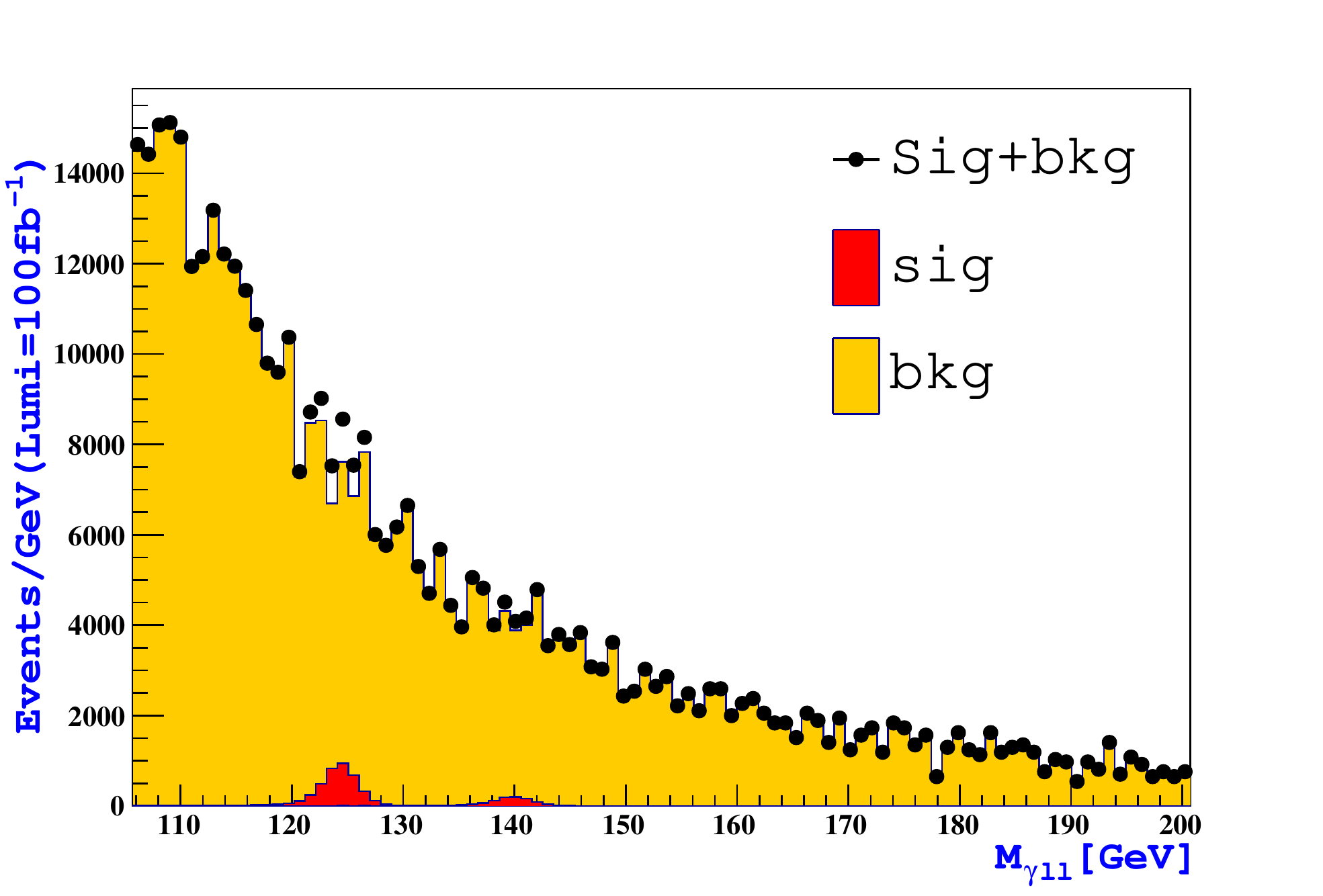}~~\includegraphics[width=7.5cm,height=6cm,angle=0]{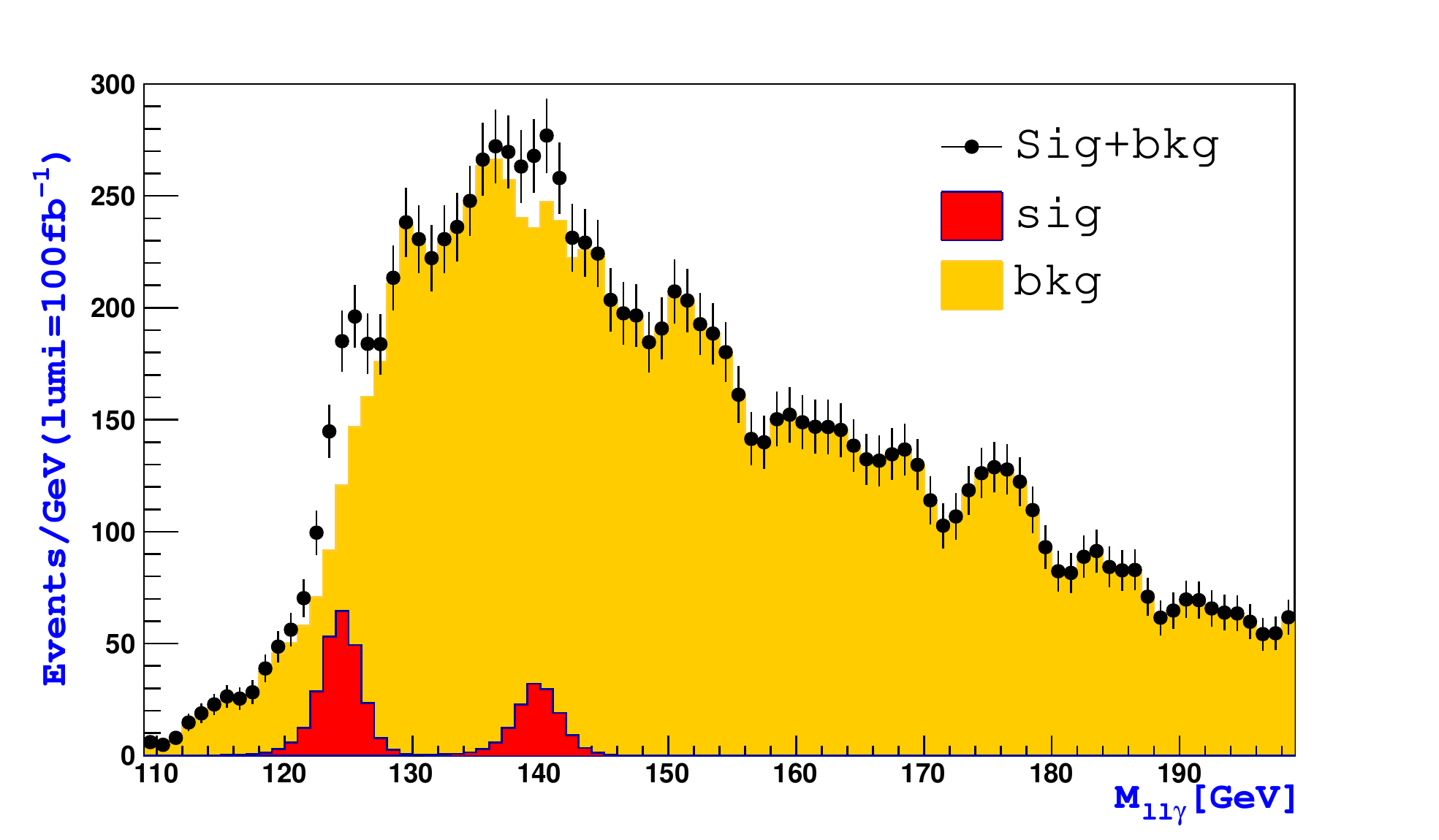}
\caption{Number of signal events for $h$ and $h^\prime \to Z(\to \ell^+\ell^-)\gamma$ decays  ($\ell=e,\mu$) (red))
induced by ggF and VBF versus the $\ell^+\ell^-\gamma$ invariant mass
at $\sqrt s=13$ TeV after 100 fb$^{-1}$ of luminosity 
alongside the total background (yellow). Their sum is also shown as data points. Left(Right): Before(After) the cuts in the text are applied.}\label{fig:8}
\end{figure}

\begin{figure}[!h]\centering
\includegraphics[height=6cm,width=16cm]{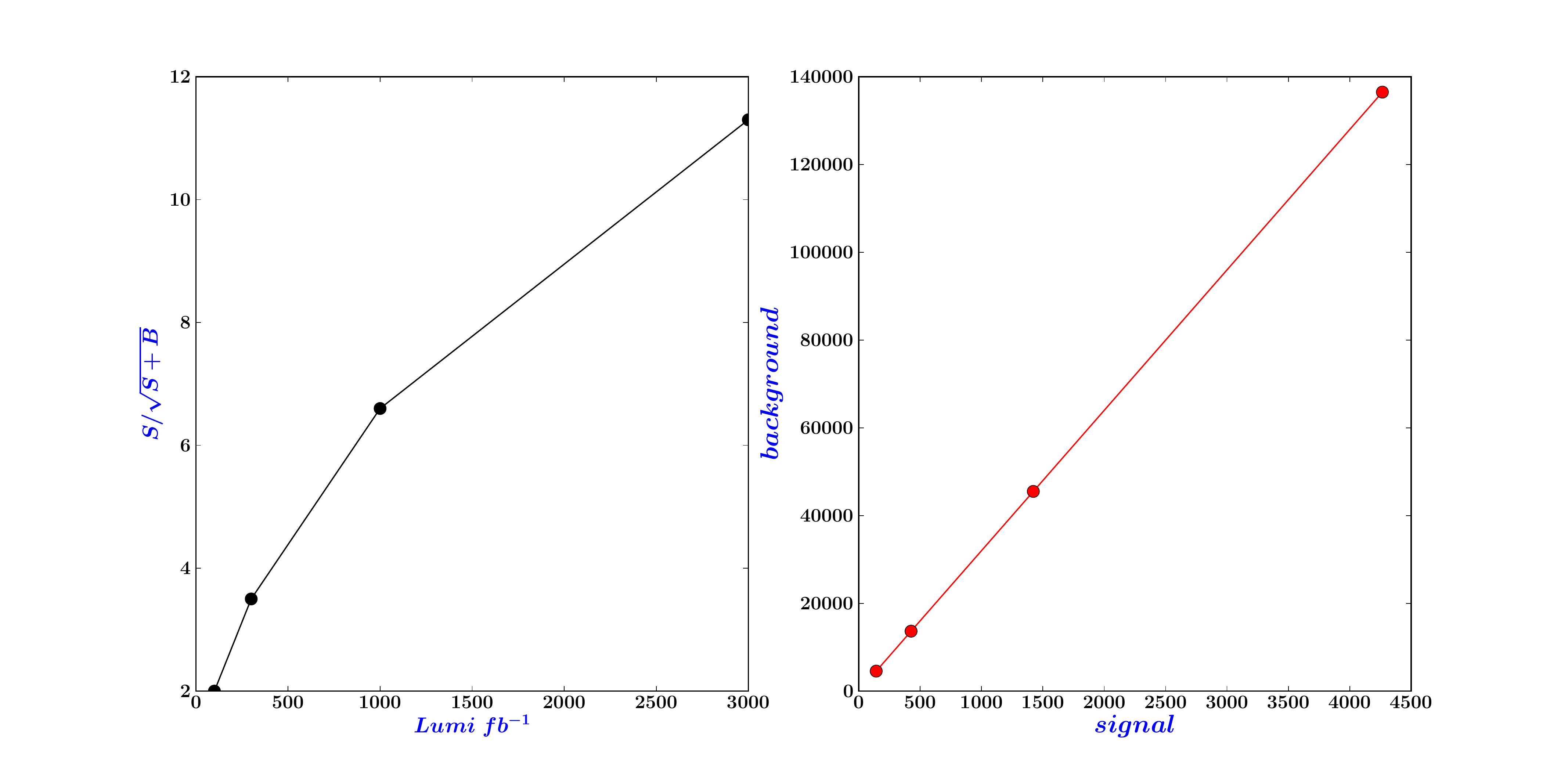}
\caption{Left: Significance of the $h^\prime \to Z(\to \ell^+\ell^-)\gamma$ signal (for $m_{h'}=140 $ GeV and $\ell=e,\mu$)
versus the luminosity (black). Right: Number of events for signal and background for variable luminosity (red). 
Data are produced 
at $\sqrt{s} = 13$ TeV and the  points correspond to an integrated luminosity of 100, 300, 1000 and 3000 fb$^{-1}$.
Notice that event rates are computed after the cuts described in the text.}\label{fig:ZA}
\end{figure}

\subsection{The $ZZ(\to 4\ell)$ decays of a light BLSSM Higgs boson}

The four leptons final state through the Higgs decay via pairs of $Z$  bosons is the most significant channel for 
Higgs detection, yet it may not be the most sensitive one to BSM effects, as its leading contribution occurs at tree level, so that mixing effects of the SM-like boson with additional Higgs boson states typically drive the BSM deviations. It was however one of the channels where an anomaly at around 140 GeV appeared following the Run 1 analyses, as intimated.  
  In the MSSM, as mentioned above, in order  to keep the signal strength of the lightest Higgs boson $h$ consistent with the observed data, one is constrained to the decoupling region, where at $M_A \gg M_Z$ and the Higgs mixing angle $\alpha \sim \beta -\frac{\pi}{2}$. Therefore, the coupling of the heaviest MSSM CP-even Higgs boson, $H$, with the SM gauge bosons is very suppressed. In the case of the BLSSM, $\tilde{g}$ plays an important role in enhancing both the first and the second lightest
CP-even Higgs boson couplings with SM gauge bosons, as discussed  in \cite{Abdallah:2014fra} and
seen in Fig.~\ref{fig:1}.

\begin{figure}[!h]\centering
\includegraphics[height=7cm,width=10cm,angle=0]{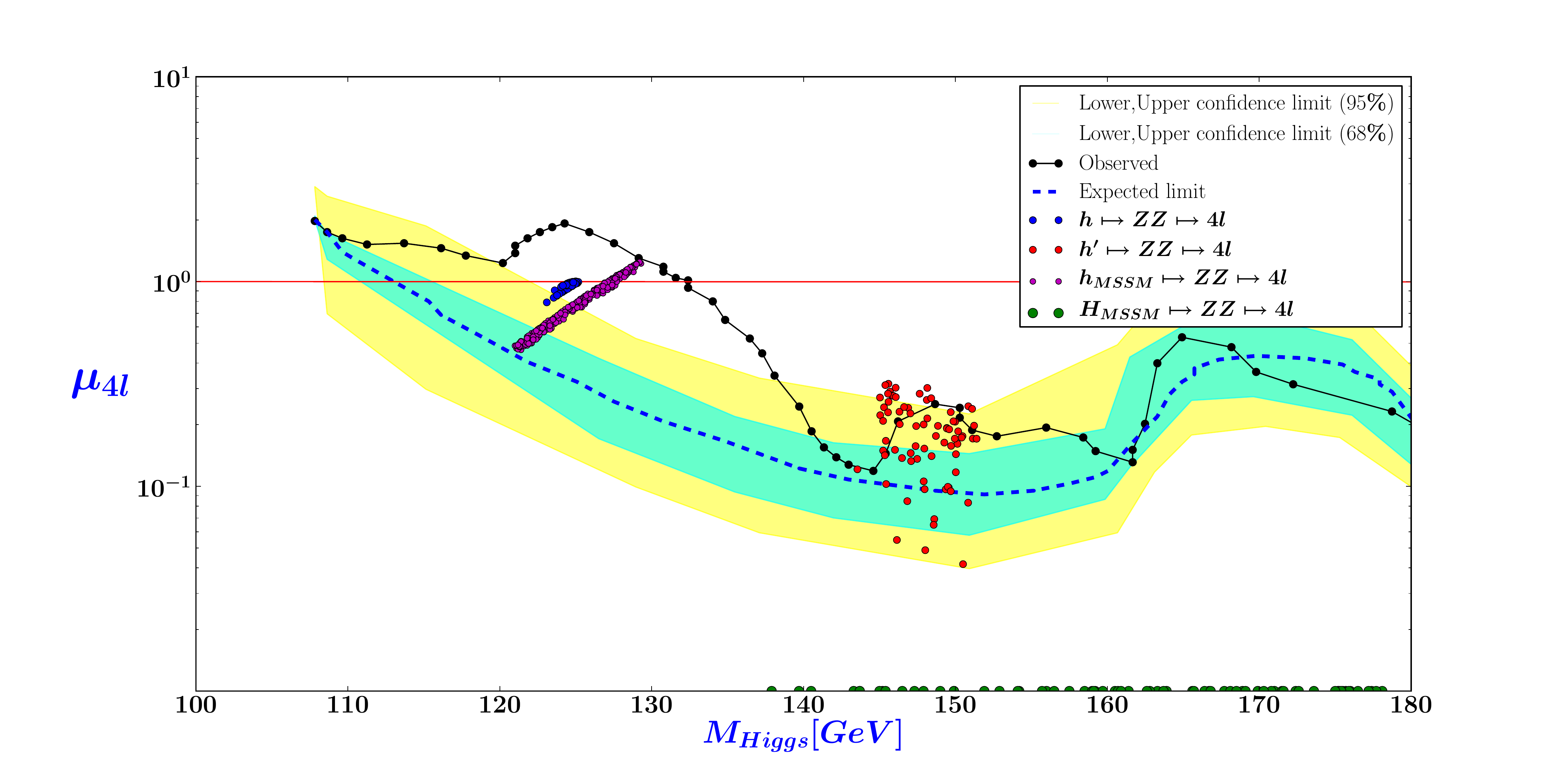}
\caption{
Signal strength of the lightest and next-to-lightest Higgs bosons in the BLSSM (in blue and red, respectively) in the $ZZ$ channel. 
The signal strength of the lightest and next-to-lightest Higgs bosons in the MSSM are given in violet and green points, respectively.
The $1$ and $2 \sigma$ confidence intervals are extracted from data collected during 
Run 1 with the observed exclusion limit as given in \cite{Chatrchyan:2013mxa} is also included.}\label{fig:muzz}
\end{figure}

In Fig. \ref{fig:muzz}, we show the signal strength of $h$ and $h'$ decays to $ZZ$ for $m_h\approx125$ GeV and 
$m_{h'}$ around $140$ GeV along with 1  and $2 \sigma$ confidence bands extracted from data collected during Run 1 with the observed exclusion limit of \cite{Aad:2014fia}. As the other two channels previously discussed, the results of the BLSSM for both $h$ and  $h'$ match the observed data rather closely. We refrain from presenting here the MSSM results for $h$ as in the 
decoupling limit they essentially coincide with the SM ones (whereas those for the MSSM  $H$ boson are outside the frame).

\begin{figure}[h]\centering
\includegraphics[width=8cm,height=6cm,angle=0]{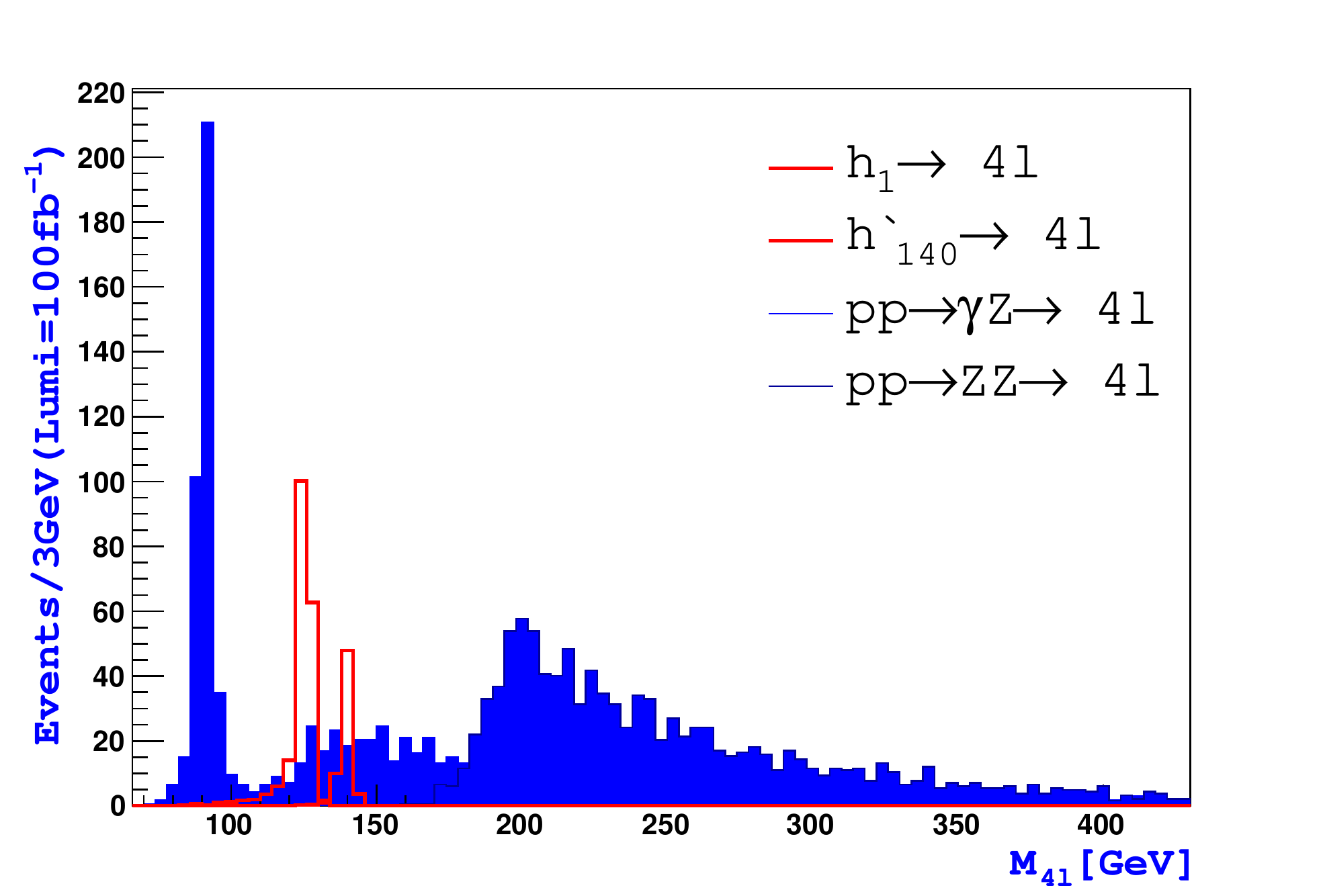}
\caption{Number of signal events for $h$ and $h^\prime \to ZZ(\to 4\ell)$ decays ($\ell=e,\mu$) (red)
induced by ggF and VBF versus the $4\ell$ invariant mass
at $\sqrt s=13$ TeV after 100 fb$^{-1}$ of luminosity 
alongside the two dominant backgrounds (blue and black). }\label{fig:9}
\end{figure}

The results of our simulation for Run 2 are based on $ZZ\to 4\ell$ decays, wherein $\ell=e,\mu$. 
In Fig.~\ref{fig:9}, we show the number of events for the $h$ and $h'$ bosons in the BLSSM plotted
against the four-lepton invariant mass. As can be seen from this plot, a promising signature of $h^\prime \to ZZ \to 4 \ell$ around $140\ {\rm GeV}$ emerges alongside the SM-like one at $\approx 125$ GeV. The main contributions from SM backgrounds come from $Z\gamma^\ast$ and $ZZ^{(*)}$. Significances at 100 fb$^{-1}$ are already enough to claim evidence
in both Higgs channels.

 Reconstruction of the $h$ and $h^\prime$ decays can only be performed for one on-shell ($Z$) and one off-shell ($Z^*$) 
gauge boson, as $M_Z<m_{h,h'}<2M_Z$ for both Higgs states. We notice that the combination of the two highest $p_T$ leptons is the most likely one to emerge from 
the on-shell $Z$ boson decay while the  other two leptons most often come from the off-shell $Z$ boson decay.
Fig.~\ref{fig:10} shows the reconstruction of both the off-shell and on-shell $Z$ boson decays for both $h$ and $h^\prime $, illustrating that the off-shell distribution can be used to increase the purity of each signal from cross-contamination.

In the light of such $Z$ boson spectra, we required the following cuts.

{\begin{enumerate}
\item The pseudorapidity of both electrons and muons is $|\eta|\le 2.5$.  
\item We require a $Z$ candidate formed with a pair of leptons of the same flavour and opposite charge, with mass window $40\le M_Z\le 120$ GeV, the remaining leptons constructing the second off-shell $Z$ boson if they satisfy $ 12\le M_Z\le 120 $ GeV. 
\item In reconstructing the on-shell $Z$ we require the highest transverse momentum lepton pair to be $ \ge 20$ GeV.
\item To protect the signals against leptons originating from hadron decays in jet fragmentation or from the decay of low-mass hadronic resonances, we require $M_{\ell^+\ell^-} \ge 4 $ GeV, where $M_{\ell^+\ell^-}$ is the invariant mass of any lepton pair. 
\end{enumerate}}

\begin{figure}[h]\centering
\includegraphics[width=7.5cm,height=6cm,angle=0]{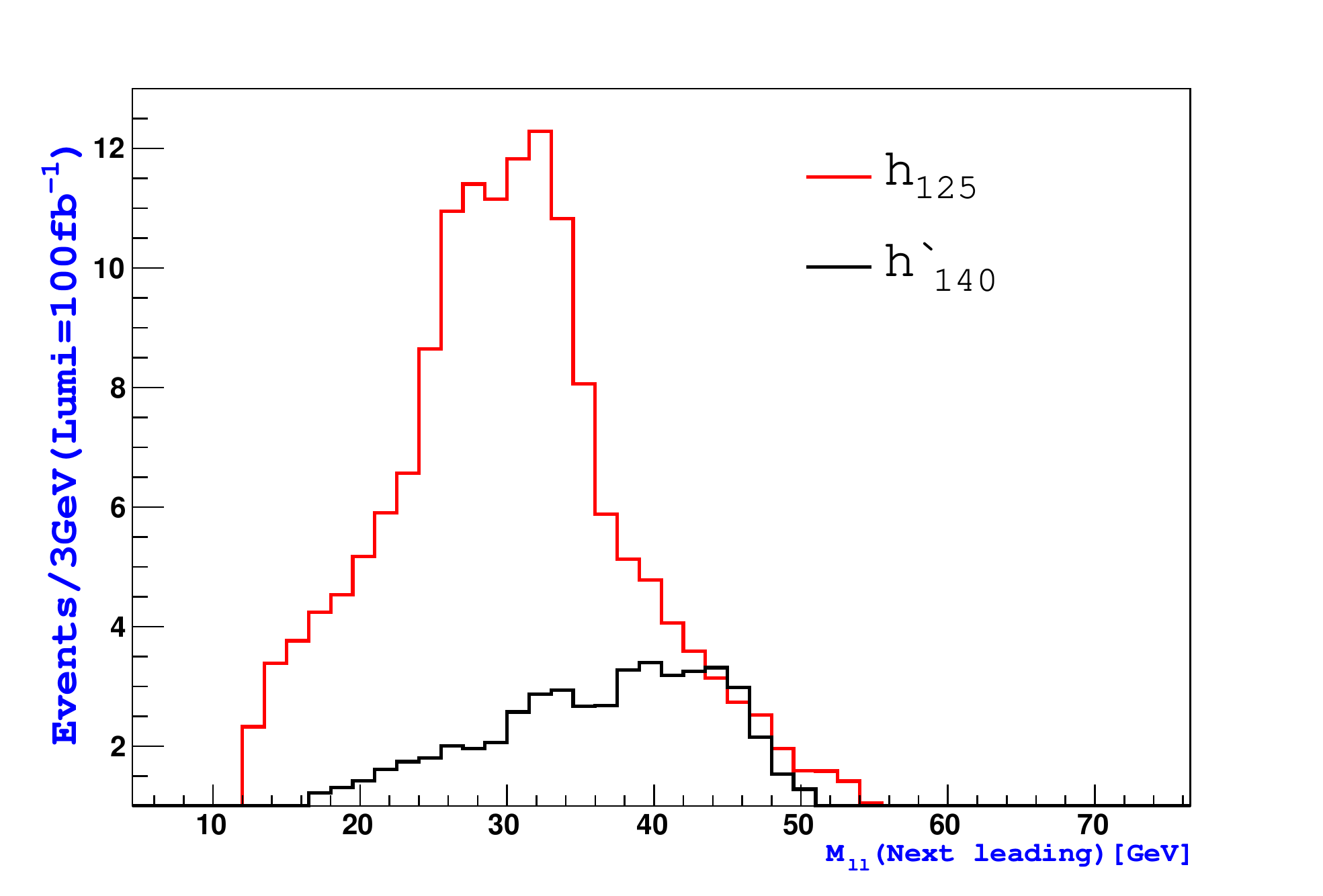}~~\includegraphics[width=7.5cm,height=6cm,angle=0]{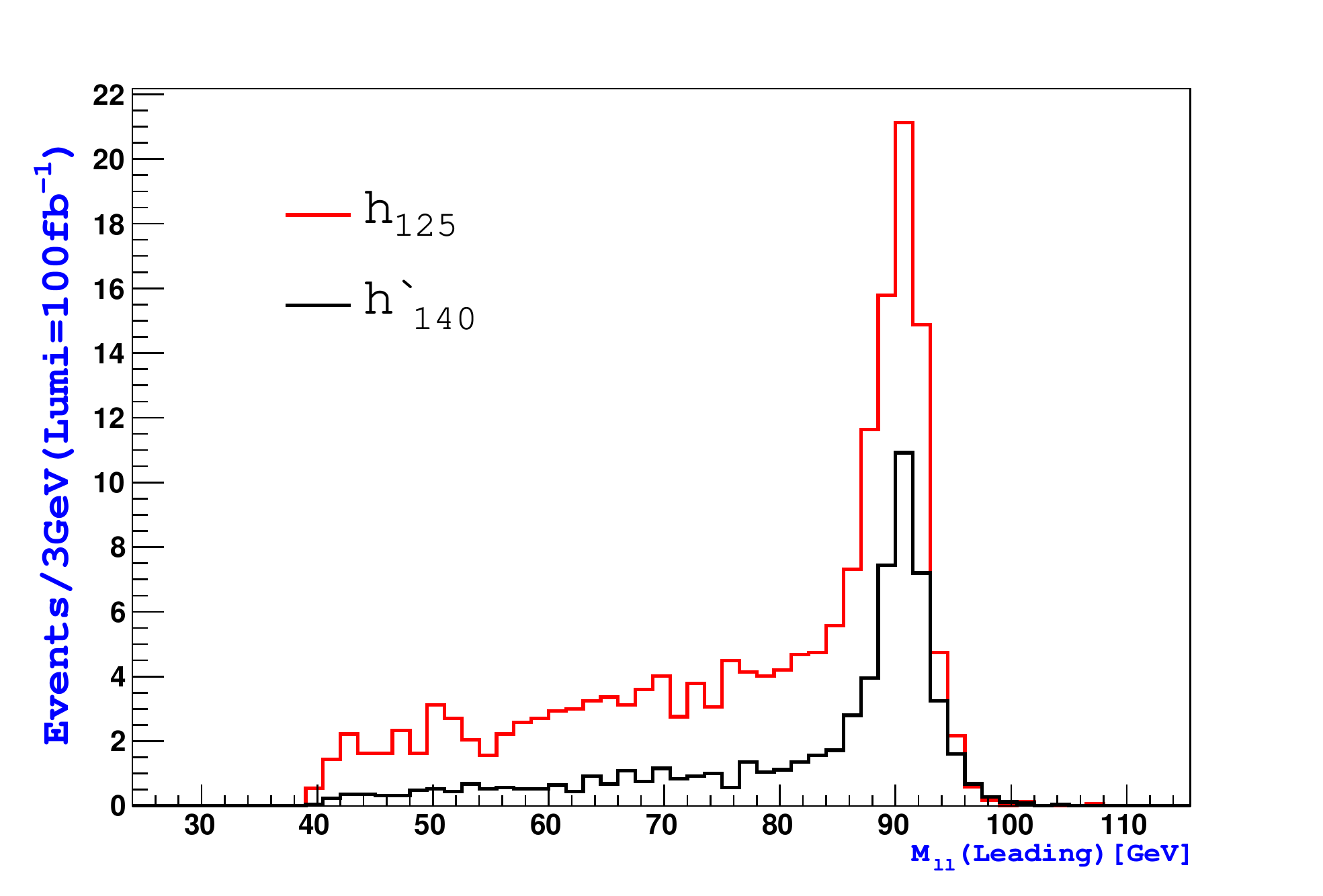}
\caption{Number of signal events for $h$ (red) and $h^\prime \to ZZ(\to 4\ell)$ (black) decays ($\ell=e,\mu$)
induced by ggF and VBF versus the $2\ell$ invariant mass
at $\sqrt s=13$ TeV after 100 fb$^{-1}$ of luminosity. Left(Right): for the off(on)-shell $Z$ case.}\label{fig:10}
\end{figure}
\noindent
Such a selection is already effective  at 100 fb$^{-1}$ and,
as usual, increasing luminosity will render this signal more and more significant, as per trend seen in Fig.~\ref{fig:11}.

\begin{figure}[!h]\centering
\includegraphics[height=6cm,width=16cm]{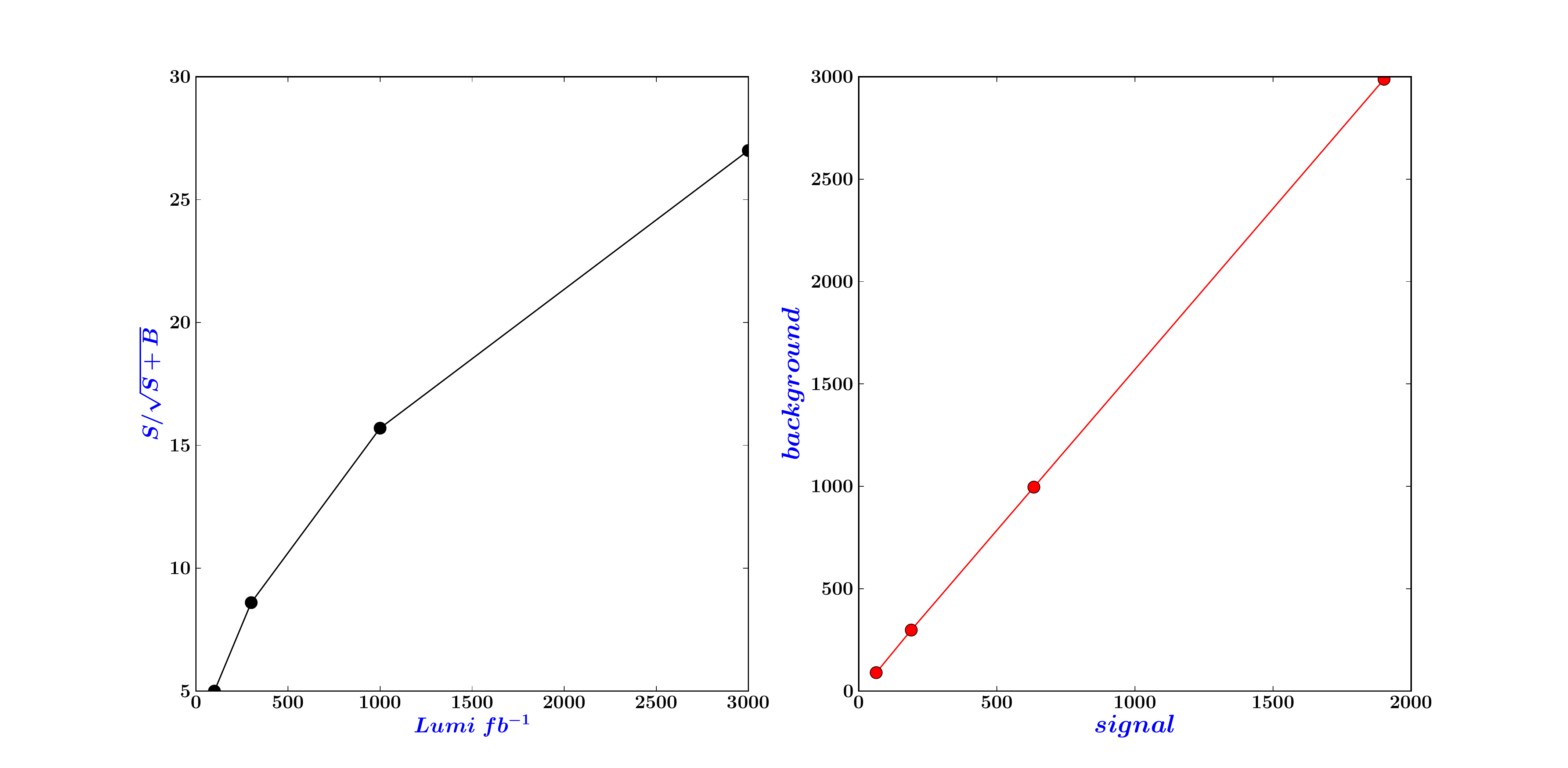}
\caption{Left: Significance of the $h^\prime \to ZZ(\to 4\ell)$ signal (for $m_{h'}=140 $ GeV and $\ell=e,\mu$)
versus the luminosity (black). Right: Number of events for signal and background for variable luminosity (red). 
Data are produced 
at $\sqrt{s} = 13$ TeV and the  points correspond to an integrated luminosity of 100, 300, 1000 and 3000 fb$^{-1}$.
Notice that event rates are computed after the cuts described in the text.}\label{fig:11}
\end{figure}


\section{Conclusion}
\label{sec:summary}

We have analysed the discovery potential of a second neutral Higgs boson in the BLSSM at the LHC. We have confirmed that a double Higgs peak structure can be accessed in this framework, in the $\gamma\gamma$, $Z(\to \ell^+\ell^-)\gamma$ and  $ZZ(\to 4\ell)$ decay channels with Higgs boson masses at  $m_h\sim 125$ GeV and $m_{h'} = 140$ GeV, wherein $h$ and $h'$ are the lightest CP-even Higgs states of the MSSM-like and genuine BLSSM spectra, respectively.

Furthermore, under the assumption that the aforementioned excesses are not confirmed by Run 2 data, we have studied the possibilities at the CERN machine of establishing signals of an heavier $h'$ state of the BLSSM. We have shown that a peculiar decay  in the BLSSM is $h'\to hh$ (i.e., into a pair of SM-like Higgs bosons), which can in fact be dominant from its threshold (at $m_{h'}\approx 2 m_h\approx 250$ GeV) onwards. We have shown that the associate $\gamma\gamma b\bar b$ signature can be spectacularly visible over a wide mass interval, from, say, 250 to 500 GeV.
 
Combining all these results, and noting that similar  Higgs signals would not be available in the MSSM, we conclude that their extraction, either around 140 GeV or anywhere beyond 250 GeV or so, would not only point to a non-minimal SUSY scenario, hence beyond the MSSM, but also possibly pinpoint  the BLSSM.

\vspace*{0.45cm}
\noindent
{\bf Acknowledgements}

\noindent
SK is partially supported by the STDF project 13858. SM is supported in part through the NExT Institute. All authors
acknowledge support from the grant H2020-MSCA-RISE-2014 n. 645722 (NonMinimalHiggs). 
AH is partially supported by the EENP2 FP7-PEOPLE-2012-IRSES grant.

\end{document}